\documentclass{CSML}
\pdfoutput=1

\usepackage{lastpage}
\newcommand{\dOi}{13(3:6)2017}
\lmcsheading{}{1--\pageref{LastPage}}{}{}%
{Dec.~10, 2016}{Jul.~21, 2017}{}

\usepackage{hyperref}
\hypersetup{hidelinks}

\usepackage{array}
\usepackage{rotating}
\usepackage{bussproofs}

\newcommand{\newrule}[3]{\newcommand{#1}[1]{\RL{#2}\csname #3IC\endcsname{##1}}}
\newcommand{\logsys}[1]{\textnormal{\textsf{#1}}}

\newcommand{\AIC}[1]{\AxiomC{\ensuremath{#1}}}
\newcommand{\UIC}[1]{\UnaryInfC{\ensuremath{#1}}}
\newcommand{\BIC}[1]{\BinaryInfC{\ensuremath{#1}}}
\newcommand{\ZIC}[1]{\AIC{}\UIC{#1}}
\newcommand{\RL}[1]{\RightLabel{\ensuremath{#1}}}
\newcommand{\DP}{\DisplayProof}
\newcommand{\ax}{\ensuremath{\textit{ax}}}
\newcommand{\cut}{\ensuremath{\textit{cut}}}
\newrule{\AX}{\ax}{Z}
\newrule{\CUT}{\cut}{B}
\newrule{\WK}{\wk}{U}
\newrule{\TOP}{\top}{Z}
\newcommand{\A}{\Gamma}
\newcommand{\B}{\Delta}
\newcommand{\ie}{\emph{i.e.}}

\newcommand{\boxit}[2]{\vspace{1ex}\newline\fbox{\begin{minipage}{0.985\textwidth}{#1 #2}\end{minipage}}\vspace{1ex}}

\newcommand{\size}[1]{\mathopen{|}#1\mathclose{|}}
\newcommand{\ssize}[1]{\sharp #1}

\newcommand{\Var}{\ensuremath{\mathcal{X}}}
\renewcommand{\O}{\ensuremath{\mathcal{O}}}
\newcommand{\FOL}{\ensuremath{\mathcal{F}}}
\newcommand{\Hexa}{\ensuremath{\mathcal{H}}}

\newcommand{\OL}{\logsys{OL}}
\newcommand{\OCL}{\logsys{OCL}\ensuremath{+}}
\newcommand{\OLf}{\OL\ensuremath{_{\textnormal{\textrm{f}}}}}
\newcommand{\pOLf}{\OL\ensuremath{_{\textnormal{\textrm{f}}}^0}}
\newcommand{\OLft}{OLf}
\newcommand{\pOLft}{OLf0}

\newcommand{\bisubst}[5]{#1\left[{^{#2}/_{#3}},{^{#4}/_{#5}}\right]}

\newcommand{\fcsign}{\mathrel{\Downarrow}}
\newcommand{\rvsign}{\mathrel{\Uparrow}}
\newcommand{\fcvdash}[2]{\vdash {#1}\fcsign{#2}}
\newcommand{\rvvdash}[2]{\vdash {#1}\rvsign{#2}}

\newcommand{\SC}[1]{\LeftLabel{[\ensuremath{#1}]}}
\newcommand{\wk}{\textit{w}}
\newcommand{\wedgerr}{{\rvsign}{\wedge}}
\newcommand{\wedgerv}{{\wedge}{\rvsign}}
\newcommand{\toprr}{{\rvsign}{\top}}
\newcommand{\toprv}{{\top}{\rvsign}}
\newcommand{\cwrr}{{\rvsign}{\textit{cw}}}
\newcommand{\cwrv}{{\textit{cw}}{\rvsign}}
\newcommand{\reacrr}{{\rvsign}{\text{R}}}
\newcommand{\reacrv}{{\text{R}}{\rvsign}}
\newcommand{\reacfc}{{\text{R}}{\fcsign}}
\newcommand{\decidl}{\text{D}_1}
\newcommand{\decidr}{\text{D}_2}
\newcommand{\axx}{\textit{ax}_v}
\newcommand{\veel}{{\vee}_1}
\newcommand{\veer}{{\vee}_2}
\newcommand{\ex}{\textit{ex}}
\newcommand{\ctr}{\textit{c}}
\newcommand{\cw}{\textit{cw}}
\newrule{\WKL}{\wk L}{U}
\newrule{\WKR}{\wk R}{U}
\newrule{\ZWEDGELl}{{\wedge}_1L}{Z}
\newrule{\ZWEDGELr}{{\wedge}_2L}{Z}
\newrule{\ZVEERl}{{\vee}_1R}{Z}
\newrule{\ZVEERr}{{\vee}_2R}{Z}
\newrule{\WEDGELl}{{\wedge}_1L}{U}
\newrule{\WEDGELr}{{\wedge}_2L}{U}
\newrule{\WEDGER}{{\wedge}R}{B}
\newrule{\VEEL}{{\vee}L}{B}
\newrule{\VEERl}{{\vee}_1R}{U}
\newrule{\VEERr}{{\vee}_2R}{U}
\newrule{\TOPR}{{\top}R}{Z}
\newrule{\BOTL}{{\bot}L}{Z}
\newrule{\NEG}{{\neg}}{U}
\newrule{\NEGNEGR}{{\neg\neg}R}{Z}
\newrule{\NEGNEGL}{{\neg\neg}L}{Z}
\newrule{\TND}{\textit{tnd}}{Z}
\newrule{\NEGR}{{\neg}R}{U}
\newrule{\NEGL}{{\neg}L}{U}
\newrule{\EX}{\ex}{U}
\newrule{\CTR}{\ctr}{U}
\newrule{\VEEl}{\veel}{U}
\newrule{\VEEr}{\veer}{U}
\newrule{\VEEi}{{\vee}_i}{U}
\newrule{\WEDGE}{{\wedge}}{B}
\newrule{\CW}{\cw}{U}
\newrule{\AXX}{\axx}{Z}
\newrule{\CWV}{\cw_{\vee}}{U}
\newrule{\WEDGErr}{\wedgerr}{B}
\newrule{\WEDGErv}{\wedgerv}{B}
\newrule{\TOPrr}{\toprr}{Z}
\newrule{\TOPrv}{\toprv}{Z}
\newrule{\REACrr}{\reacrr}{U}
\newrule{\REACrv}{\reacrv}{U}
\newrule{\CWl}{\textit{cw}_1}{U}
\newrule{\CWr}{\textit{cw}_2}{U}
\newrule{\CWrr}{\cwrr}{U}
\newrule{\CWrv}{\cwrv}{U}
\newrule{\DECID}{\text{D}}{U}
\newrule{\DECIDl}{\decidl}{U}
\newrule{\DECIDr}{\decidr}{U}
\newrule{\REACfc}{\reacfc}{U}

\newcommand{\cutxa}{\ensuremath{v\textit{-cut}_1}}
\newcommand{\cutxb}{\ensuremath{v\textit{-cut}_2}}
\newcommand{\cutxc}{\ensuremath{v\textit{-cut}_3}}
\newcommand{\cuta}{\ensuremath{\textit{cut}_1}}
\newcommand{\cutb}{\ensuremath{\textit{cut}_2}}
\newcommand{\cutc}{\ensuremath{\textit{cut}_3}}
\newcommand{\cutd}{\ensuremath{\textit{cut}_4}}
\newcommand{\cute}{\ensuremath{\textit{cut}_5}}
\newcommand{\cutra}{\ensuremath{\textit{cut}_0}}
\newcommand{\cutrb}{\ensuremath{\textit{cut}'_0}}
\newrule{\CUTXA}{\cutxa}{B}
\newrule{\CUTXB}{\cutxb}{B}
\newrule{\CUTXC}{\cutxc}{B}
\newrule{\CUTA}{\cuta}{B}
\newrule{\CUTB}{\cutb}{B}
\newrule{\CUTC}{\cutc}{B}
\newrule{\CUTD}{\cutd}{B}
\newrule{\CUTE}{\cute}{B}
\newrule{\CUTRA}{\cutra}{B}
\newrule{\CUTRB}{\cutrb}{B}

\newcommand{\fboundsym}{\varphi}
\newcommand{\sboundsym}{\psi}
\newcommand{\fbound}[1]{\fboundsym(#1)}
\newcommand{\sbound}[1]{\sboundsym(#1)}

\newcommand{\cunea}{\texttt{E}_1}
\newcommand{\cuneb}{\texttt{E}_2}
\newcommand{\cunec}{\texttt{E}_3}

\newcommand{\cf}{\logsys{cf}}
\newcommand{\fw}{\logsys{fw}}
\newcommand{\bwf}{\logsys{bw}\ensuremath{_{\textnormal{\textrm{f}}}}}
\newcommand{\fwf}{\logsys{fw}\ensuremath{_{\textnormal{\textrm{f}}}}}
\newcommand{\coq}{\texttt{\upshape Coq}}
\newcommand{\ocaml}{\texttt{\upshape OCaml}}

\newcommand{\timo}{\textsf{\textsc{to}}}

\begin{document}

\title{Focusing in Orthologic}
\author{Olivier Laurent}
\address{Universit\'e de Lyon, CNRS, ENS de Lyon, Universit\'e Claude Bernard Lyon 1, LIP}
\email{olivier.laurent@ens-lyon.fr}
\thanks{This work was supported by the LABEX MILYON (ANR-10-LABX-0070) of Universit\'e de Lyon, within the program ``Investissements d'Avenir'' (ANR-11-IDEX-0007), and by projects R\'ecr\'e (ANR-11-BS02-0010) and Elica (ANR-14-CE25-0005), all operated by the French National Research Agency (ANR)}

\titlecomment{This paper is an extended version of~\cite{olf}.}

\begin{abstract}
  We propose new sequent calculus systems for \emph{orthologic} (also known as \emph{minimal quantum logic}) which satisfy the cut elimination property. The first one is a simple system relying on the involutive status of negation. The second one incorporates the notion of \emph{focusing} (coming from linear logic) to add constraints on proofs and to optimise proof search.
We demonstrate how to take benefits from the new systems in automatic proof search for orthologic.
\end{abstract}

\maketitle

Classical (propositional) logic can be used to reason about facts in classical mechanics and is related with the lattice structure of Boolean algebras. On its side, quantum (propositional) logic has been introduced to represent observable facts in quantum mechanics. It is provided as an axiomatization of the lattice structure of the closed subspaces of Hilbert spaces.
This corresponds to the structure of so-called orthomodular lattices. Among the properties of these lattices, and thus of quantum logic, one finds the orthomodularity law ($p\leq q \Longrightarrow q\leq p \vee (\neg p \wedge q)$) which is a very weak form of distributivity. Removing this law gives the notion of \emph{ortholattice} and leads to the associated \emph{orthologic} (also called minimal quantum logic, as it can be defined as quantum logic without orthomodularity).
The interested reader can find more about logic and quantum physics in~\cite{lqp}.

In the description and reasoning about quantum properties, quantum logic is more accurate than orthologic. Nevertheless a formula valid in orthologic is also valid in quantum logic, and thus provides a valid quantum property.
In the current state of the art, orthologic benefits of much better logical properties than plain quantum logic (in proof theory in particular)~\cite{semortho,cutelimlattices,orthomodnotelem,ptmql,gentzenql}. It moreover corresponds to an interesting class of lattices: ortholattices, widely studied in lattice theory. Ortholattices are bounded lattices with an involutive negation such that $p\vee \neg p = \top$. As a consequence they can be understood as Boolean lattices without distributivity, and indeed distributive ortholattices are exactly Boolean lattices.

The main topic of the present work is the study of the proof theory of orthologic, from the sequent calculus point of view. Sound and complete sequent calculi satisfying the cut-elimination property already occur in the literature (see for example~\cite{cutelimlattices,ptmql,blqlcutelim}). Our first result is another such calculus which is particularly simple: each sequent has exactly two formulas and only seven rules are required. It relies on ideas of W.~Tait~\cite{normalderiv} for the representation of systems with an involutive negation (also promoted by J.-Y.~Girard in linear logic~\cite{ll}), and shows how orthologic can be seen as an extension of the additive fragment of linear logic with one new contraction-weakening rule.
The second and main contribution of this paper lies in the development of a ``second-level proof-theory'' for orthologic by investigating the notion of focusing in this setting.

Focusing, introduced in linear logic by J.-M.~Andreoli~\cite{focusing}, is a constraint on the structure of proofs which requires connectives sharing some structural properties (like reversibility) to be grouped together. The key point is that this restriction is sound and complete: focused proofs are proofs and any provable sequent admits a focused proof. Together with cut elimination, focusing can be used as a strong tool in proof search and proof study since it reduces the search space to focused proofs.
Focusing has also been used to define new logical systems~\cite{lc}.

In the case of orthologic, we show that focusing can be defined, and interacts particularly well with the two-formulas sequents. In particular, not only logical rules associated with connectives are constrained but also structural rules can be organised. The exchange rule can be hidden easily in the specific focusing rules and the contraction-weakening rule becomes precisely constrained. As a consequence, we obtain a bound on the height of all focused proofs of a given sequent, which is rarely the case in the presence of a contraction rule.
Starting from this remark, we experiment proof search strategies for orthologic based on our focused system.

\bigskip

In Section~\ref{secortholat}, we recall the definition of ortholattice and orthologic with the main results from the literature on sequent calculus and cut-elimination for orthologic. In Section~\ref{secol}, we introduce the sequent calculus \OL{} (inspired by additive linear logic) with a few properties. Section~\ref{secfol} gives the two-steps construction of the focused system \OLf. We explain how focusing is applied to orthologic and we prove soundness, completeness and cut-elimination. The last Section~\ref{secps} is dedicated to the application of \OLf{} in (backward and forward) proof search for orthologic. This is based on upper bounds on the height of proofs and on additional structural properties of focused cut-free proofs.

\bigskip
Most of the results of the paper have been formalised in the \coq{} proof assistant and naive versions of the proof search algorithms are implemented in \ocaml{} (see page~\pageref{pageaddmat}).

\section{Ortholattices and Orthologic}\label{secortholat}

\emph{Orthologic} or \emph{minimal quantum logic} is the logic associated with the order relation of \emph{ortholattices} (for some results about ortholattices, see for example~\cite{birkhofflattices}).

\begin{defi}[Ortholattice]
An \emph{ortholattice} $\O$ is a bounded lattice (a lattice with smallest and biggest elements $\bot$ and $\top$) with an order-reversing involution $p\mapsto \neg p$ (also often denoted $p^\bot$ in the literature), called \emph{orthocomplement}, satisfying $p\vee \neg p = \top$ (for all $p$ in $\O$).
\end{defi}

In particular the following properties hold for any two elements $p$ and $q$ of any ortholattice:
\begin{align*}
  p \leq q &\Longrightarrow \neg q\leq \neg p\\
  \neg \neg p &= p \\
  \neg \bot &= \top \\
  \neg (p \vee q) &= \neg p \wedge \neg q \\
  p \wedge \neg p &= \bot
\end{align*}
as well as the other De Morgan's laws, but there is no distributivity law between $\wedge$ and $\vee$.

\begin{exa}[Hexagon Ortholattice]\label{exhexa}
The \emph{hexagon lattice} \Hexa{} below (also called benzene ring) is an ortholattice:
  
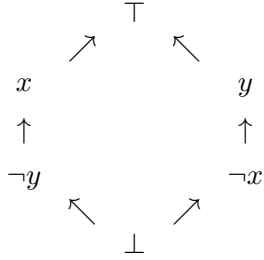
\begin{equation*}
    \begin{array}{ccccc}
      & & \top \\
      & \nearrow & & \nwarrow \\
      x & & & & y \\[1ex]
      \uparrow & & & & \uparrow \\[1ex]
      \neg y & & & & \neg x \\
      & \nwarrow & & \nearrow \\
      & & \bot
    \end{array}
  \end{equation*}
Orthomodularity, which can be stated as the equation $\forall p q,\,p\vee q = p\vee(\neg p\wedge(p\vee q))$, does not hold in \Hexa{} since:
\begin{align*}
  \neg y \vee x &= x \\
  \neg y\vee(\neg \neg y\wedge(\neg y\vee x)) &= \neg y\vee (y\wedge x)= \neg y
\end{align*}
In fact, any non-orthomodular ortholattice contains \Hexa{} as a sub-ortholattice (see for example~\cite[Theorem~5.4]{beranoml}).
This means this lattice is typical of what may happen in orthologic but not in quantum logic (see also Proposition~\ref{propimppb}).
\end{exa}

\emph{Orthologic} is the logic associated with the class of ortholattices, or conversely ortholattices are the algebras associated with orthologic.
Formulas in orthologic are built using connectives corresponding to the basic operations of ortholattices:
\begin{equation*}
  A ::= X \mid A \wedge A \mid A \vee A \mid \top \mid \bot \mid \neg A
\end{equation*}
where $X$ ranges over elements of a given countable set $\Var$ of \emph{variables}.

We want then $A\vdash B$ to be derivable in orthologic if and only if $A\leq B$ is true in any ortholattice $\O$ (for every interpretation of variables as elements of $\O$, and with connectives in $A$ and $B$ interpreted through the corresponding operations of $\O$).
In particular, the Lindenbaum algebra associated with orthologic over the set $\Var$ is the free ortholattice $\FOL$ over $\Var$ (which is infinite as soon as $\Var$ contains at least two elements~\cite{freeortho}).

If we adopt a sequent calculus style presentation, an (sound and complete) axiomatization of orthologic can be given by the following axioms and rules (in the spirit of~\cite{semortho}):
\begin{gather*}
  \AX{A\vdash A}
  \DP
\qquad\qquad
  \AIC{A\vdash B}
  \AIC{B\vdash C}
  \CUT{A\vdash C}
  \DP
\\[2ex]
  \ZWEDGELl{A\wedge B\vdash A}
  \DP
\qquad
  \ZWEDGELr{A\wedge B\vdash B}
  \DP
\qquad\quad
  \AIC{C\vdash A}
  \AIC{C\vdash B}
  \WEDGER{C\vdash A\wedge B}
  \DP
\qquad\quad
  \TOPR{C\vdash \top}
  \DP
\\[2ex]
  \ZVEERl{A\vdash A\vee B}
  \DP
\qquad
  \ZVEERr{B\vdash A\vee B}
  \DP
\qquad\quad
  \AIC{A\vdash C}
  \AIC{B\vdash C}
  \VEEL{A\vee B \vdash C}
  \DP
\qquad\quad
  \BOTL{\bot\vdash C}
  \DP
\\[2ex]
  \AIC{A\vdash B}
  \NEG{\neg B\vdash \neg A}
  \DP
\qquad\quad
  \NEGNEGR{A\vdash \neg\neg A}
  \DP
\qquad\quad
  \NEGNEGL{\neg\neg A\vdash A}
  \DP
\qquad\quad
  \TND{\top\vdash A\vee\neg A}
  \DP
\end{gather*}
The first line corresponds to an (pre) order relation. The second and third lines correspond to a bounded inf semi-lattice and bounded sup semi-lattice (thus together they provide us the structure of a bounded lattice). The fourth line adds the missing ortholattice ingredients related with the orthocomplement $\neg A$.

\begin{exa}\label{exone}
If one wants to prove that for any $p$ and $q$ in an ortholattice, we have:
\begin{equation*}
\top\leq ((p\wedge q)\vee \neg p)\vee \neg q
\end{equation*}
We can either use algebraic properties of ortholattices (which have to be proved as well):
\begin{align*}
    ((p\wedge q)\vee \neg p)\vee \neg q
& = (p\wedge q)\vee (\neg p \vee \neg q) \\
& = (p\wedge q)\vee \neg (p \wedge q) \\
& = \top
\end{align*}
or we can use, on the logic side, a derivation with conclusion the corresponding sequent $\top\vdash ((X\wedge Y)\vee \neg X)\vee \neg Y$. This requires us to use most of the rules above (see landscape figure on page~\pageref{pagefig}).
\begin{sidewaysfigure}\label{pagefig}
\vspace{15cm}
\centering
\scalebox{0.74}{
  \ZVEERl{\neg X\vdash \neg X\vee \neg Y}
  \NEG{\neg(\neg X\vee \neg Y)\vdash \neg\neg X}
  \NEGNEGL{\neg\neg X\vdash X}
  \CUT{\neg(\neg X\vee \neg Y)\vdash X}
  \ZVEERr{\neg Y\vdash \neg X\vee \neg Y}
  \NEG{\neg(\neg X\vee \neg Y)\vdash \neg\neg Y}
  \NEGNEGL{\neg\neg Y\vdash Y}
  \CUT{\neg(\neg X\vee \neg Y)\vdash Y}
  \WEDGER{\neg(\neg X\vee \neg Y)\vdash X\wedge Y}
  \NEG{\neg (X\wedge Y)\vdash \neg\neg(\neg X\vee \neg Y)}
  \NEGNEGL{\neg\neg(\neg X\vee \neg Y)\vdash \neg X\vee \neg Y}
  \CUT{\neg (X\wedge Y)\vdash \neg X\vee \neg Y}
  \ZVEERr{\neg X\vee \neg Y\vdash (X\wedge Y)\vee (\neg X\vee \neg Y)}
  \CUT{\neg (X\wedge Y)\vdash (X\wedge Y)\vee (\neg X\vee \neg Y)}
  \noLine
  \UIC{(1)}
  \DP
}
\\[6ex]
\scalebox{0.74}{
  \ZVEERr{\neg X\vdash (X\wedge Y)\vee \neg X}
  \ZVEERl{(X\wedge Y)\vee \neg X\vdash ((X\wedge Y)\vee \neg X)\vee \neg Y}
  \CUT{\neg X\vdash ((X\wedge Y)\vee \neg X)\vee \neg Y}
  \ZVEERr{\neg Y\vdash ((X\wedge Y)\vee \neg X)\vee \neg Y}
  \VEEL{\neg X\vee \neg Y\vdash ((X\wedge Y)\vee \neg X)\vee \neg Y}  
  \noLine
  \UIC{(2)}
  \DP
}
\\[6ex]
\scalebox{0.74}{
  \ZVEERl{X\wedge Y\vdash (X\wedge Y)\vee \neg X}
  \ZVEERl{(X\wedge Y)\vee \neg X\vdash ((X\wedge Y)\vee \neg X)\vee \neg Y}
  \CUT{X\wedge Y\vdash ((X\wedge Y)\vee \neg X)\vee \neg Y}
  \AIC{(2)}
  \noLine
  \UIC{\neg X\vee \neg Y\vdash ((X\wedge Y)\vee \neg X)\vee \neg Y}  
  \VEEL{(X\wedge Y)\vee (\neg X\vee \neg Y)\vdash ((X\wedge Y)\vee \neg X)\vee \neg Y}
  \noLine
  \UIC{(3)}
  \DP
}
\\[6ex]
\scalebox{0.74}{
  \TND{\top\vdash (X\wedge Y)\vee \neg (X\wedge Y)}
  \ZVEERl{X\wedge Y\vdash (X\wedge Y)\vee (\neg X\vee \neg Y)}
  \AIC{(1)}
  \noLine
  \UIC{\neg (X\wedge Y)\vdash (X\wedge Y)\vee (\neg X\vee \neg Y)}
  \VEEL{(X\wedge Y)\vee \neg (X\wedge Y)\vdash (X\wedge Y)\vee (\neg X\vee \neg Y)}
  \CUT{\top\vdash (X\wedge Y)\vee (\neg X\vee \neg Y)}
  \AIC{(3)}
  \noLine
  \UIC{(X\wedge Y)\vee (\neg X\vee \neg Y)\vdash ((X\wedge Y)\vee \neg X)\vee \neg Y}
  \CUT{\top\vdash ((X\wedge Y)\vee \neg X)\vee \neg Y}
  \DP
}
\end{sidewaysfigure}
\end{exa}

The axiomatization proposed above is a direct translation of the order-theoretic definition of ortholattices. From a proof-theoretic point of view, it has strong defects such as the impossibility of eliminating the cut rule:
\begin{prooftree}
  \AIC{A\vdash B}
  \AIC{B\vdash C}
  \CUT{A\vdash C}
\end{prooftree}
(which encodes the transitivity of the order relation).
Example~\ref{exone} could not be derived without this rule for example.
A reason for trying to avoid the cut rule is that when studying a property like $A\vdash C$, the cut rule tells us that we may need to invent some arbitrary $B$ (unrelated with $A$ and $C$). This may lead us to difficulties, undecidability, etc. In the opposite, cut-free systems usually satisfy the \emph{sub-formula property} stating that every formula appearing in a proof of a given sequent is a sub-formula of a formula of this sequent.
The idea of finding presentations of the logic associated with lattices in such a way that cut (or transitivity) could be eliminated goes back to P.~Whitman~\cite{freelattices} with applications to the theory of lattices.
In the case of ortholattices, one can find such an axiomatization in~\cite{cutelimlattices} under the name \OCL{} (also called \logsys{GOL} in~\cite{proofsearchortho}):
\boxit{\OCL}{
\begin{gather*}
  \AX{A\vdash A}
  \DP
\qquad\qquad
  \AIC{\A\vdash \B}
  \WKL{\A,A\vdash \B}
  \DP
\qquad
  \AIC{\A\vdash \B}
  \WKR{\A\vdash A,\B}
  \DP
\\[2ex]
  \AIC{\A,A\vdash \B}
  \WEDGELl{\A,A\wedge B\vdash \B}
  \DP
\qquad
  \AIC{\A,B\vdash \B}
  \WEDGELr{\A,A\wedge B\vdash \B}
  \DP
\qquad\qquad
  \AIC{\A\vdash A,\B}
  \AIC{\A\vdash B,\B}
  \WEDGER{\A\vdash A\wedge B,\B}
  \DP
\\[2ex]
  \AIC{\A\vdash A,\B}
  \VEERl{\A\vdash A\vee B,\B}
  \DP
\qquad
  \AIC{\A\vdash B,\B}
  \VEERr{\A\vdash A\vee B,\B}
  \DP
\qquad\qquad
  \AIC{\A,A\vdash \B}
  \AIC{\A,B\vdash \B}
  \VEEL{\A,A\vee B\vdash \B}
  \DP
\\[2ex]
  \TOPR{\A\vdash \top,\B}
  \DP
\qquad
  \BOTL{\A,\bot\vdash \B}
  \DP
\qquad\qquad
  \AIC{\A,A\vdash \B}
  \NEGR{\A\vdash \neg A,\B}
  \DP
\qquad
  \AIC{\A\vdash A,\B}
  \NEGL{\A,\neg A\vdash \B}
  \DP
\end{gather*}
where sequents $\A\vdash\B$ are given from two finite \emph{sets} $\A$ and $\B$ of formulas such that $\ssize{\A}+\ssize{\B}\leq 2$ (comma denotes set union and $\ssize{}$ the cardinality of a set).
}

\begin{exa}
  We can prove in \OCL{} the sequent of Example~\ref{exone}:
  \begin{prooftree}
    \AX{X\vdash X}
    \NEGR{\vdash X,\neg X}
    \VEERr{\vdash X,(X\wedge Y)\vee \neg X}
    \VEERl{\vdash X,((X\wedge Y)\vee \neg X)\vee \neg Y}
    \AX{Y\vdash Y}
    \NEGR{\vdash Y,\neg Y}
    \VEERr{\vdash Y,((X\wedge Y)\vee \neg X)\vee \neg Y}
    \WEDGER{\vdash X\wedge Y,((X\wedge Y)\vee \neg X)\vee \neg Y}
    \VEERl{\vdash (X\wedge Y)\vee \neg X,((X\wedge Y)\vee \neg X)\vee \neg Y}
    \VEERl{\vdash ((X\wedge Y)\vee \neg X)\vee \neg Y}
    \WKL{\top\vdash ((X\wedge Y)\vee \neg X)\vee \neg Y}
  \end{prooftree}
\end{exa}

The following key properties of \OCL{} are proved in~\cite{cutelimlattices}:

\begin{thm}[Cut Elimination in \OCL]\label{thmcutelimocl}
  The cut rule
  \AIC{\A_1\vdash A,\B_1}
  \AIC{\A_2,A\vdash \B_2}
  \BIC{\A_1,\A_2\vdash\B_1,\B_2}
  \DP
is admissible in \OCL.
\qed
\end{thm}

\begin{thm}[Soundness and Completeness of \OCL]
  \OCL{} is sound and complete for orthologic.
\qed
\end{thm}

By looking at the structure of the rules, one can see there is an important symmetry between $\vee$ on the left and $\wedge$ on the right, $\wedge$ on the left and $\vee$ on the right, $\bot$ on the left and $\top$ on the right, etc. This is not very surprising in a context where negation is an involution, and this is an incarnation of De Morgan's duality between $\wedge$ and $\vee$ and $\top$ and $\bot$. W.~Tait~\cite{normalderiv} (followed by J.-Y.~Girard in linear logic~\cite{ll}) has shown how to simplify sequent calculi in the presence of an involutive negation by restricting negation to variables and by considering one-sided sequents only.
This idea has been partly applied in~\cite{blqlcutelim} where they define formulas for orthologic as:
\begin{equation*}
  A ::= X \mid A \wedge A \mid A \vee A \mid \top \mid \bot \mid \neg X
\end{equation*}
and negation is then extended to all formulas by induction (it is not a true connective anymore):
\begin{gather*}
  \neg (\neg X) := X \quad
  \neg (\bot) := \top \quad
  \neg (\top) := \bot \quad
  \neg (A\vee B) := \neg A\wedge \neg B \quad
  \neg (A\wedge B) := \neg A\vee \neg B
\end{gather*}
so that we obtain $\neg\neg A = A$ for any A. However the system proposed in~\cite{blqlcutelim} does not really take benefits from this encoded involutive negation on formulas, since they use two-sided sequents. One can also note that no remark is given in~\cite{blqlcutelim} regarding the number of formulas in sequents. However one can see that, in their system, $\A\vdash \B$ is provable if and only if $\bigwedge\A\vdash \bigvee\B$ is provable, and that a proof of a sequent $\A\vdash \B$ with at most one formula in $\A$ and at most one formula in $\B$ contains only sequents satisfying this property.

We propose to go further in this direction of involutive negation to target a simpler sequent calculus system for orthologic.

\section{One-Sided Orthologic}\label{secol}

In order to clarify the analysis and to be closer to an implementation, we prefer to consider sequents based on lists rather than sets or multi-sets. The main difference with respect to \OCL{} is the necessity to use an explicit contraction rule and an explicit exchange rule.
We thus consider two kinds of sequents: ${}\vdash A,B$ and ${}\vdash A$. As a notation, $\Pi$ corresponds to $0$ or $1$ formula so that ${}\vdash A,\Pi$ is a common notation for both kinds of sequents.
Like in~\cite{blqlcutelim}, formulas are built with negation on variables only:
\begin{equation*}
  A ::= X \mid A \wedge A \mid A \vee A \mid \top \mid \bot \mid \neg X
\end{equation*}
and, by moving to a one-sided list-based system, the derivation rules we obtain are:
\begin{gather*}
  \AX{\vdash \neg A,A}
  \DP
\qquad\qquad
  \AIC{\vdash A,B}
  \EX{\vdash B,A}
  \DP
\qquad\qquad
  \AIC{\vdash A,A}
  \CTR{\vdash A}
  \DP
\qquad\qquad
  \AIC{\vdash A}
  \WK{\vdash A,B}
  \DP
\\[2ex]
  \AIC{\vdash A,\Pi}
  \VEEl{\vdash A\vee B,\Pi}
  \DP
\qquad
  \AIC{\vdash B,\Pi}
  \VEEr{\vdash A\vee B,\Pi}
  \DP
\qquad\qquad
  \AIC{\vdash A,\Pi}
  \AIC{\vdash B,\Pi}
  \WEDGE{\vdash A\wedge B,\Pi}
  \DP
\qquad\qquad
  \TOP{\vdash \top,\Pi}
  \DP
\end{gather*}
Note, a version with sequents as multi-sets of formulas with at most 2 elements would simply lead us to the discarding of the exchange rule.

We are going to optimise these one-sided rules in order to build our new system \OL. We give here an informal description of the path leading to the new set of rules. First, we can assume $\Pi$ not to be empty in the rules above since the case of an empty $\Pi$ is derivable from the non-empty case (using rules ($\ctr$) and ($\wk$)). For example, for the ($\veel$) rule:
\begin{prooftree}
  \AIC{\vdash A}
  \WK{\vdash A,A\vee B}
  \VEEl{\vdash A\vee B,A\vee B}
  \CTR{\vdash A\vee B}
\end{prooftree}
Second, once we thus consider only logical rules with two formulas in sequents, the only rule with a premise with only one formula is the ($\wk$) rule and the only rule with a conclusion with only one formula is the ($\ctr$) rule. This means that in a proof of a sequent with two formulas, ($\ctr$) and ($\wk$) rules always come together, one above the other, and we can group them into a new combined rule:
\begin{equation*}
  \AIC{\vdash A,A}
  \CTR{\vdash A}
  \WK{\vdash A,B}
  \DP
\qquad\qquad\mapsto\qquad\qquad
  \AIC{\vdash A,A}
  \CW{\vdash A,B}
  \DP
\end{equation*}
Finally a sequent ${}\vdash A$ can always be encoded as ${}\vdash A,A$ since one is provable if and only if the other is (thanks to the rules ($\ctr$) and ($\wk$)).
We thus focus on sequents ${}\vdash A,B$ only, and on the following set of rules:
\boxit{\OL}{
\begin{gather*}
  \AX{\vdash \neg A,A}
  \DP
\qquad\qquad
  \AIC{\vdash A,B}
  \EX{\vdash B,A}
  \DP
\qquad\qquad
  \AIC{\vdash A,A}
  \CW{\vdash A,B}
  \DP
\\[2ex]
  \AIC{\vdash A,C}
  \VEEl{\vdash A\vee B,C}
  \DP
\qquad
  \AIC{\vdash B,C}
  \VEEr{\vdash A\vee B,C}
  \DP
\qquad\qquad
  \AIC{\vdash A,C}
  \AIC{\vdash B,C}
  \WEDGE{\vdash A\wedge B,C}
  \DP
\qquad
  \TOP{\vdash \top,C}
  \DP
\end{gather*}
}\newline
This sequent calculus with $7$ rules ($6$ rules in its multi-set-based and set-based versions) does not seem to occur in the literature and looks simpler than all the sound and complete calculi for orthologic we have found. We call it \OL.
Relying on the remarks above, we have:
\begin{thm}[Soundness and Completeness of \OL]\label{thmolsndcmp}
  ${}\vdash \neg A,B$ is provable in \OL{} if and only if $A\vdash B$ is provable in \OCL{}, so that \OL{} is sound and complete for orthologic.
\end{thm}

\proof
To be completely precise, we have to recall that formulas of \OL{} are all formulas of \OCL. While the converse is not true, there is a canonical mapping of formulas of \OCL{} into formulas of \OL{} obtained by unfolding the definition of $\neg$.
Both implications are obtained by induction on proofs.

For soundness, we rely on Theorem~\ref{thmcutelimocl}. For example, in the case of a ($\cw$) rule:
\begin{equation*}
  \AIC{\vdash\neg A,\neg A}
  \CW{\vdash\neg A,B}
  \DP
\qquad\mapsto\qquad
  \AX{A\vdash A}
  \NEGL{A,\neg A\vdash}
  \NEGR{A\vdash\neg\neg A}
  \AIC{A\vdash \neg A}
  \NEGL{A,\neg\neg A\vdash}
  \CUT{A\vdash}
  \WKR{A\vdash B}
  \DP
\end{equation*}

Concerning completeness, we prove simultaneously that:
\begin{center}
  \begin{tabular}{r@{ }c@{ }l}
   $A\vdash B$ & in \OCL{} entails & ${}\vdash \neg A,B$ in \OL \\
   $A\vdash {}$ & in \OCL{} entails & ${}\vdash \neg A,\neg A$ in \OL{} \\
   $A,A'\vdash {}$ & in \OCL{} entails & ${}\vdash \neg A,\neg A'$ and ${}\vdash \neg A',\neg A$ in \OL{} (for $A\neq A'$)\\
   ${}\vdash B$ & in \OCL{} entails & ${}\vdash B,B$ in \OL \\
   ${}\vdash B,B'$ & in \OCL{} entails & ${}\vdash B,B'$ and ${}\vdash B',B$ in \OL{} (for $B\neq B'$)
  \end{tabular}
\end{center}
For example:
\begin{equation*}
  \AIC{A\wedge B,B\vdash}
  \WEDGELr{A\wedge B\vdash}
  \WKR{A\wedge B\vdash C}
  \DP
\qquad\mapsto\qquad
  \AIC{\vdash \neg B,\neg A\vee\neg B}
  \VEEr{\vdash \neg A\vee\neg B,\neg A\vee\neg B}
  \CW{\vdash \neg A\vee\neg B,C}
  \DP \tag*{\raisebox{-1.25em}{\qEd}}
\end{equation*}
\def\popQED{}

For readers familiar with linear logic~\cite{ll}, this calculus \OL{} can be seen as one-sided additive linear logic extended with the ($\cw$) rule, if we replace $\vee$ by $\oplus$, $\wedge$ by $\mathbin{\&}$ and $\bot$ by $0$.

\begin{exa}
  We can prove in \OL{} the sequent of Example~\ref{exone} in its one-sided version:
  \begin{prooftree}
    \AX{\vdash \neg X,X}
    \VEEr{\vdash (X\wedge Y)\vee \neg X,X}
    \VEEl{\vdash ((X\wedge Y)\vee \neg X)\vee \neg Y,X}
    \EX{\vdash X,((X\wedge Y)\vee \neg X)\vee \neg Y}
    \AX{\vdash \neg Y,Y}
    \VEEr{\vdash ((X\wedge Y)\vee \neg X)\vee \neg Y,Y}
    \EX{\vdash Y,((X\wedge Y)\vee \neg X)\vee \neg Y}
    \WEDGE{\vdash X\wedge Y,((X\wedge Y)\vee \neg X)\vee \neg Y}
    \VEEl{\vdash (X\wedge Y)\vee \neg X,((X\wedge Y)\vee \neg X)\vee \neg Y}
    \VEEl{\vdash ((X\wedge Y)\vee \neg X)\vee \neg Y,((X\wedge Y)\vee \neg X)\vee \neg Y}
    \CW{\vdash ((X\wedge Y)\vee \neg X)\vee \neg Y,\bot}
    \EX{\vdash \bot,((X\wedge Y)\vee \neg X)\vee \neg Y}
  \end{prooftree}
\end{exa}

We now describe a few properties of \OL{} which will be used later.

First, the cut rule
  \AIC{\vdash A,B}
  \AIC{\vdash \neg B,C}
  \BIC{\vdash A,C}
  \DP
\;
is admissible. It is possible to give a direct proof of this result but this will happen here as a consequence of a stronger result we have to prove later anyway (see Proposition~\ref{propolcutelim}). It can be deduced from Theorems~\ref{thmolsndcmp} and~\ref{thmcutelimocl} as well.

We also have by simple inductions:
\begin{prop}[Axiom expansion for \OL]\label{propoletaexp}
  If we restrict the axiom rule of \OL{} to its variable case
  \AXX{\vdash \neg X,X}\DP,
the general rule ($\ax$) is derivable.
\qed
\end{prop}

\begin{lem}[Reversibility of $\wedge$]\label{lemolrevw}
  ${}\vdash A\wedge B,C$ is provable iff both ${}\vdash A,C$ and ${}\vdash B,C$ are.
\qed
\end{lem}

\begin{lem}[Reversing]\label{lemolrenv}
  If we restrict the ($\cw$) rule to formulas of the shape $A_1\vee A_2$:
  \begin{prooftree}
    \AIC{\vdash A_1\vee A_2,A_1\vee A_2}
    \CWV{\vdash A_1\vee A_2,B}
  \end{prooftree}
where moreover $B$ is neither $\top$ nor a $\wedge$, the general rule ($\cw$) is admissible.
\end{lem}

\proof
This is done in two steps, first by proving the restriction on $A$ (by induction on $A$ for an arbitrary $B$) and then the restriction on $B$ (by induction on $B$, with $A=A_1\vee A_2$).
\qed

This means that restricting the contraction-weakening rule of \OL{} to $\vee$-formulas does not modify the expressiveness of the system.

\section{Focused Orthologic}\label{secfol}

Relying on the strong relation between the sequent calculus \OL{} and linear logic, we import the idea of \emph{focusing}~\cite{focusing}.
This constraint on the structure of proofs is based on an analysis of the polarity of connectives, by separating those which are reversible and those which are not. By reducing the space of proofs of each formula, it is a strong tool for accelerating proof search.
In orthologic, the connectives $\wedge$ and $\top$ are reversible: the conclusion of their introduction rule implies its premises (see Lemma~\ref{lemolrevw} for example).
Such connectives are also called \emph{asynchronous} or negative. Their dual connectives are called \emph{synchronous} or positive.
Following this pattern, we separate formulas into synchronous and asynchronous ones according to their main connective:
\begin{center}
\begin{tabular}{r@{}l}
 $X$, $\bot$ and $A\vee B$ & {} are synchronous,\\
 and $\neg X$, $\top$ and $A\wedge B$ & {} are asynchronous.
\end{tabular}
\end{center}
So that $A$ is synchronous (resp.\ asynchronous) if and only if $\neg A$ is asynchronous (resp.\ synchronous). The choice for variables is in fact arbitrary, as soon as we preserve this dual polarity between $X$ and $\neg X$ for each of them.

Let us now apply focusing to orthologic and to \OL{} in particular.

\subsection{A First Focused System \texorpdfstring{\pOLf}{\OLft}}

Dealing with variables in focused systems is delicate, so we recommend the reader not very familiar with focusing to concentrate on the other aspects of the system first.

A key result will be to prove the focused system to be as expressive as \OL{} (and thus sound and complete for orthologic). In order to make this as simple and clear as possible, we will work in two steps. Indeed some optimisations (to be introduced later on in Section~\ref{secolf}) would make a direct translation more difficult.

Our first focused system \pOLf{} is based on four kinds of sequents. For each of them, we give an \emph{informal explanation} based on how we can find a proof of such a sequent, thus from the point of view of a bottom-up reading of proofs and rules:
\begin{itemize}
\item In a sequent ${}\rvvdash{}{A,B}$, all the asynchronous connectives at the roots of $A$ and $B$ (in formulas $A$ and $B$ seen as trees) will be deconstructed and after that, $A$ and $B$ are turned into some $A'$ and $B'$ which are synchronous (or negation of a variable) and allowed to move to the left of $\rvsign$. In fact we first work on $A$ and then we move to a sequent ${}\rvvdash{A'}{B}$ (with $A'$ synchronous or negation of a variable), and we start working on $B$.
\item In a sequent ${}\rvvdash{A}{B}$, $A$ is synchronous or is the negation of a variable. The asynchronous connectives at the root of $B$ will be deconstructed and after that we reach some ${}\rvvdash{A}{B'}$ with $B'$ synchronous (or negation of a variable) and allowed to move to the left of $\rvsign$.
\item In a sequent ${}\rvvdash{A,B}{}$, $A$ and $B$ are synchronous or the negation of a variable. We have to select a synchronous formula among them (let say $B$) and start decomposing its synchronous connectives at the root, in a sequent ${}\fcvdash{A}{B}$. Before that, we can apply contraction-weakening rules to $A$ and $B$. This is the main place where choices have to be made during proof search.
\item In a sequent ${}\fcvdash{A}{B}$, $A$ is synchronous or is the negation of a variable. The synchronous connectives at the root of $B$ will be deconstructed and after that we reach some ${}\fcvdash{A}{B'}$ with $B'$ asynchronous (and we will start decomposing its asynchronous connectives at the root in a sequent ${}\rvvdash{A}{B'}$). Choices concerning the decomposition of $\vee$ will have to be made here.
\end{itemize}
Note, sequents ${}\rvvdash{}{A,B}$ are crucial for the comparison with other systems but play a weak role inside this system. Indeed they occur only in proofs of sequents of the same shape and only at the bottom part of such a proof. As soon as we reach a sequent ${}\rvvdash{\_}{\_}$ (in the bottom-up reading of a proof), we will not find any other sequent ${}\rvvdash{}{\_,\_}$ above.

Let us be more formal now with the explicit list of the rules of the system \pOLf{} which uses four kinds of sequents ${}\rvvdash{}{A,B}$, ${}\rvvdash{A}{B}$, ${}\rvvdash{A,B}{}$ and ${}\fcvdash{A}{B}$:
\boxit{\pOLf}{
\begin{gather*}
  \AIC{\rvvdash{}{A,C}}
  \AIC{\rvvdash{}{B,C}}
  \WEDGErr{\rvvdash{}{A\wedge B,C}}
  \DP
\qquad
  \TOPrr{\rvvdash{}{\top,C}}
  \DP  
\qquad
  \AIC{\rvvdash{A}{C}}
  \REACrr{\rvvdash{}{A,C}}
  \DP  
\\[2ex]
  \AIC{\rvvdash{C}{A}}
  \AIC{\rvvdash{C}{B}}
  \WEDGErv{\rvvdash{C}{A\wedge B}}
  \DP
\qquad
  \SC{\text{(s) or (n)}}
  \TOPrv{\rvvdash{A}{\top}}
  \DP  
\qquad
  \AIC{\rvvdash{C,A}{}}
  \REACrv{\rvvdash{C}{A}}
  \DP  
\\[2ex]
  \AIC{\rvvdash{C,C}{}}
  \SC{\text{(s) or (n)}}
  \CWl{\rvvdash{C,A}{}}
  \DP
\qquad\qquad
  \AIC{\rvvdash{C,C}{}}
  \SC{\text{(s) or (n)}}
  \CWr{\rvvdash{A,C}{}}
  \DP
\\[2ex]
  \AIC{\fcvdash{C}{A}}
  \SC{\text{(s)}}
  \DECIDl{\rvvdash{A,C}{}}
  \DP
\qquad
  \AIC{\fcvdash{C}{A}}
  \SC{\text{(s)}}
  \DECIDr{\rvvdash{C,A}{}}
  \DP
\\[2ex]
  \AXX{\fcvdash{\neg X}{X}}
  \DP
\qquad
  \AIC{\fcvdash{C}{A}}
  \VEEl{\fcvdash{C}{A\vee B}}
  \DP
\qquad
  \AIC{\fcvdash{C}{B}}
  \VEEr{\fcvdash{C}{A\vee B}}
  \DP
\qquad
  \AIC{\rvvdash{C}{A}}
  \SC{\text{(a)}}
  \REACfc{\fcvdash{C}{A}}
  \DP
\end{gather*}
with the following side conditions written between square brackets $[\_]$:

\hfil
$\text{(a) $A$ is asynchronous} \qquad
\text{(s) $A$ is synchronous} \qquad
\text{(n) $A$ is the negation of a variable}$.
}

One could have been more explicit by asking $[\text{(s) or (n)}]$ as side condition in the ($\reacrr$) and ($\reacrv$) rules but the following lemma proves these two side conditions to be redundant.

\begin{lem}\label{lemsidecond}
  If ${}\rvvdash{A}{C}$ or ${}\rvvdash{A,B}{}$ or ${}\fcvdash{A}{C}$ is provable then $A$ and $B$ are synchronous or the negation of a variable.
\qed
\end{lem}

\begin{exa}\label{exuniq}
  The sequent ${}\vdash (X\vee A)\vee B,(C\vee (D\vee\neg X))\wedge\top$ has many proofs in the systems of the previous sections, in particular in \OL. However the corresponding sequent ${}\rvvdash{}{(X\vee A)\vee B,(C\vee (D\vee\neg X))\wedge\top}$ has a unique proof in \pOLf:
  \begin{prooftree}
    \AXX{\fcvdash{\neg X}{X}}
    \VEEl{\fcvdash{\neg X}{X\vee A}}
    \VEEl{\fcvdash{\neg X}{(X\vee A)\vee B}}
    \DECIDl{\rvvdash{(X\vee A)\vee B,\neg X}{}}
    \REACrv{\rvvdash{(X\vee A)\vee B}{\neg X}}
    \REACfc{\fcvdash{(X\vee A)\vee B}{\neg X}}
    \VEEr{\fcvdash{(X\vee A)\vee B}{D\vee\neg X}}
    \VEEr{\fcvdash{(X\vee A)\vee B}{C\vee (D\vee\neg X)}}
    \DECIDr{\rvvdash{(X\vee A)\vee B,C\vee (D\vee\neg X)}{}}
    \REACrv{\rvvdash{(X\vee A)\vee B}{C\vee (D\vee\neg X)}}
    \TOPrv{\rvvdash{(X\vee A)\vee B}{\top}}
    \WEDGErv{\rvvdash{(X\vee A)\vee B}{(C\vee (D\vee\neg X))\wedge\top}}
    \REACrr{\rvvdash{}{(X\vee A)\vee B,(C\vee (D\vee\neg X))\wedge\top}}
  \end{prooftree}
This shows how focusing adds constraints to the structure of proofs.
\end{exa}

\begin{exa}
  We can prove the sequent corresponding to Example~\ref{exone}:
  \begin{prooftree}
    \AXX{\fcvdash{\neg X}{X}}
    \DECIDl{\rvvdash{X,\neg X}{}}
    \REACrv{\rvvdash{X}{\neg X}}
    \REACfc{\fcvdash{X}{\neg X}}
    \VEEr{\fcvdash{X}{(X\wedge Y)\vee \neg X}}
    \VEEl{\fcvdash{X}{((X\wedge Y)\vee \neg X)\vee \neg Y}}
    \DECIDl{\rvvdash{((X\wedge Y)\vee \neg X)\vee \neg Y,X}{}}
    \REACrv{\rvvdash{((X\wedge Y)\vee \neg X)\vee \neg Y}{X}}
    \AXX{\fcvdash{\neg Y}{Y}}
    \DECIDl{\rvvdash{Y,\neg Y}{}}
    \REACrv{\rvvdash{Y}{\neg Y}}
    \REACfc{\fcvdash{Y}{\neg Y}}
    \VEEr{\fcvdash{Y}{((X\wedge Y)\vee \neg X)\vee \neg Y}}
    \DECIDl{\rvvdash{((X\wedge Y)\vee \neg X)\vee \neg Y,Y}{}}
    \REACrv{\rvvdash{((X\wedge Y)\vee \neg X)\vee \neg Y}{Y}}
    \WEDGErv{\rvvdash{((X\wedge Y)\vee \neg X)\vee \neg Y}{X\wedge Y}}
    \REACfc{\fcvdash{((X\wedge Y)\vee \neg X)\vee \neg Y}{X\wedge Y}}
    \VEEl{\fcvdash{((X\wedge Y)\vee \neg X)\vee \neg Y}{(X\wedge Y)\vee \neg X}}
    \VEEl{\fcvdash{((X\wedge Y)\vee \neg X)\vee \neg Y}{((X\wedge Y)\vee \neg X)\vee \neg Y}}
    \DECIDr{\rvvdash{((X\wedge Y)\vee \neg X)\vee \neg Y,((X\wedge Y)\vee \neg X)\vee \neg Y}{}}
    \CWr{\rvvdash{\bot,((X\wedge Y)\vee \neg X)\vee \neg Y}{}}
    \REACrv{\rvvdash{\bot}{((X\wedge Y)\vee \neg X)\vee \neg Y}}
    \REACrr{\rvvdash{}{\bot,((X\wedge Y)\vee \neg X)\vee \neg Y}}
  \end{prooftree}
\end{exa}

One can prove the soundness of \pOLf{} with respect to orthologic by translation into \OL.

\begin{prop}[Soundness of \pOLf]\label{proppolftool}
  If $\rvvdash{}{A,B}$ or $\rvvdash{A}{B}$ or $\rvvdash{A,B}{}$ or $\fcvdash{A}{B}$ is provable in \pOLf{} then $\vdash A,B$ is provable in \OL.
\end{prop}

\proof
  By a simple induction on the proof by erasing $\rvsign$ and $\fcsign$ in sequents and thanks to exchange rules in \OL.
\qed

To conclude this section, here are a few simple facts which will be useful later:
\begin{lem}\label{lembifoc}\hfill
\begin{itemize}
\item ${}\rvvdash{X,Y}{}$, ${}\rvvdash{\neg X,\neg Y}{}$ and ${}\rvvdash{\bot,\bot}{}$ are not provable (both if $X=Y$ or $X\neq Y$);
\item if ${}\rvvdash{A,B}{}$ is provable then ${}\rvvdash{B,A}{}$ as well (and with a proof of the same size);
\item if ${}\rvvdash{A,A}{}$ is provable then $A$ is synchronous and the proof contains a proof of ${}\fcvdash{A}{A}$.
\end{itemize} 
\end{lem}

\proof
Simple inductions on proofs.
\qed

\subsection{Cut Elimination in \texorpdfstring{\pOLf}{\pOLft}}

Due to the very rigid structure of proofs in focused systems, the possibility of enriching them with admissible cut rules is often used in their study~\cite{lc,focalllp} (in particular for expressiveness analysis). It is the tool we are going to use here in order to prove the completeness of \pOLf{} with respect to orthologic.

\begin{thm}[Cut Elimination in \pOLf]\label{thmpolfcutelim}
The following cut rules are admissible in \pOLf:
\begin{gather*}
  \AIC{\rvvdash{A,X}{}}
  \AIC{\rvvdash{C,\neg X}{}}
  \CUTXA{\rvvdash{A,C}{}}
  \DP
\\[2ex]
\begin{array}{c}
  \text{$C$ synchronous}
  \\[1ex]
  \AIC{\fcvdash{X}{A}}
  \AIC{\fcvdash{\neg X}{C}}
  \CUTXB{\fcvdash{C}{A}}
  \DP
\end{array}
\qquad\quad
\begin{array}{c}
  \text{$C$ synchronous}
  \\[1ex]
  \AIC{\rvvdash{X}{A}}
  \AIC{\fcvdash{\neg X}{C}}
  \CUTXC{\rvvdash{C}{A}}
  \DP
\end{array}
\\[2ex]
\displaybreak[0]
\begin{array}{c}
  \text{$B$ asynchronous or variable}
  \\[1ex]
  \AIC{\rvvdash{A}{B}}
  \AIC{\rvvdash{C}{\neg B}}
  \CUTA{\rvvdash{A,C}{}}
  \DP
\end{array}
\\[2ex]
\displaybreak[0]
  \AIC{\rvvdash{A}{B}}
  \AIC{\rvvdash{C,\neg B}{}}
  \CUTB{\rvvdash{A,C}{}}
  \DP
\qquad\qquad
\begin{array}{c}
  \text{$B$ asynchronous}
  \\[1ex]
  \AIC{\rvvdash{A}{B}}
  \AIC{\fcvdash{\neg B}{C}}
  \CUTC{\fcvdash{A}{C}}
  \DP
\end{array}
\\[2ex]
\displaybreak[0]
\begin{array}{c}
  \text{$B$ asynchronous or variable}
  \\[1ex]
  \AIC{\rvvdash{A}{B}}
  \AIC{\fcvdash{C}{\neg B}}
  \CUTD{\rvvdash{A,C}{}}
  \DP
\end{array}
\qquad\qquad
  \AIC{\rvvdash{A}{B}}
  \AIC{\rvvdash{\neg B}{C}}
  \CUTE{\rvvdash{A}{C}}
  \DP
\\[2ex]
\displaybreak[0]
  \AIC{\rvvdash{}{A,B}}
  \AIC{\rvvdash{}{C,\neg B}}
  \CUTRA{\rvvdash{}{A,C}}
  \DP
\qquad\qquad
  \AIC{\rvvdash{}{A,B}}
  \AIC{\rvvdash{C}{\neg B}}
  \CUTRB{\rvvdash{C}{A}}
  \DP
\end{gather*}
\end{thm}

\proof
This is a proof involving many cases which require a precise management of the four kinds of sequents. We try to explain the key ingredients which work in successive steps.
\begin{itemize}
\item We prove simultaneously the admissibility of ($\cutxb$) and ($\cutxc$) by induction on the size of the left premise.
\item We deduce the admissibility of ($\cutxa$) by induction on the size of the left premise.
For example:
\begin{equation*}
  \AIC{\fcvdash{X}{A}}
  \DECIDl{\rvvdash{A,X}{}}
  \AIC{\fcvdash{\neg X}{C}}
  \DECIDl{\rvvdash{C,\neg X}{}}
  \CUTXA{\rvvdash{A,C}{}}
  \DP
\quad\rightsquigarrow\quad
  \AIC{\fcvdash{X}{A}}
  \AIC{\fcvdash{\neg X}{C}}
  \dashedLine
  \CUTXB{\fcvdash{C}{A}}
  \DECIDl{\rvvdash{A,C}{}}
  \DP
\end{equation*}
since $A$ and $C$ are synchronous.
\item Using the previous steps, we prove simultaneously the admissibility of ($\cuta$), ($\cutb$), ($\cutc$), ($\cutd$) and ($\cute$) by induction on the pair $(f,p)$ where $f$ is the size of the cut-formula $B$ and $p$ is the size of the right premise.
The crucial cases are the following ones:
\begin{itemize}
\item Starting from:
  \begin{prooftree}
    \AIC{\rvvdash{A}{B_1}}
    \AIC{\rvvdash{A}{B_2}}
    \WEDGErv{\rvvdash{A}{B_1\wedge B_2}}
    \AIC{\fcvdash{C}{\neg B_1}}
    \VEEl{\fcvdash{C}{\neg B_1\vee\neg B_2}}
    \CUTD{\rvvdash{A,C}{}}
  \end{prooftree}
we can apply the induction hypothesis with a smaller cut formula by means of ($\cuta$) with Lemma~\ref{lembifoc} or ($\cutd$):
\begin{equation*}
\begin{array}{c}
  \text{$B_1$ synchronous}
  \\[1ex]
  \AIC{\rvvdash{C}{\neg B_1}}
  \AIC{\rvvdash{A}{B_1}}
  \dashedLine
  \CUTA{\rvvdash{C,A}{}}
  \dashedLine
  \UIC{\rvvdash{A,C}{}}
  \DP
\end{array}
\qquad\qquad
\begin{array}{c}
  \text{$B_1$ asynchronous}
  \\[1ex]
  \AIC{\rvvdash{A}{B_1}}
  \AIC{\fcvdash{C}{\neg B_1}}
  \dashedLine
  \CUTD{\rvvdash{A,C}{}}
  \DP
\end{array}
\end{equation*}
\item In the following case:
\begin{prooftree}
  \AIC{\rvvdash{A}{B}}
  \AIC{\fcvdash{\neg B}{C}}
  \DECIDl{\rvvdash{C,\neg B}{}}
  \CUTB{\rvvdash{A,C}{}}
\end{prooftree}
if $B$ is asynchronous, we have:
\begin{equation*}
  \AIC{\rvvdash{A}{B}}
  \AIC{\fcvdash{\neg B}{C}}
  \DECIDl{\rvvdash{C,\neg B}{}}
  \CUTB{\rvvdash{A,C}{}}
  \DP
\quad\rightsquigarrow\quad
  \AIC{\rvvdash{A}{B}}
  \AIC{\fcvdash{\neg B}{C}}
  \dashedLine
  \CUTC{\fcvdash{A}{C}}
  \DECIDr{\rvvdash{A,C}{}}
  \DP
\end{equation*}
otherwise $B$ is a variable so that $\rvvdash{A}{X}$ must come from ($\reacrv$) and we apply ($\cutxa$).
\item The most tricky case is contraction where we need two induction steps (we use here Lemma~\ref{lembifoc}):
\begin{equation*}
    \AIC{\rvvdash{A}{B}}
    \AIC{\fcvdash{\neg B}{\neg B}}
    \DECID{\rvvdash{\neg B,\neg B}{}}
    \doubleLine
    \CW{\rvvdash{\neg B,\neg B}{}}
    \CWr{\rvvdash{C,\neg B}{}}
    \CUTB{\rvvdash{A,C}{}}
    \DP
\quad\rightsquigarrow\quad
    \AIC{\rvvdash{A}{B}}
    \AIC{\rvvdash{A}{B}}
    \AIC{\fcvdash{\neg B}{\neg B}}
    \dashedLine
    \CUTC{\fcvdash{A}{\neg B}}
    \dashedLine
    \CUTD{\rvvdash{A,A}{}}
    \CWl{\rvvdash{A,C}{}}
    \DP
\end{equation*}
First we apply ($\cutc$) with a smaller right premise and then, by transforming one more step the ($\cutd$), we reach a smaller cut formula.
\end{itemize}
\item We deduce the case ($\cutrb$) and then ($\cutra$), by induction on the size of the left premise.
\qed
\end{itemize}

Among the 10 cut rules considered in the theorem above, mainly two will be used now (namely $\cutra$ and $\cutrb$). The other rules were however necessary as intermediary steps to prove the admissibility of these two rules.

\subsection{Completeness of \texorpdfstring{\pOLf}{\pOLft}}

We are going to translate proofs of \OL{} into proofs of \pOLf. We start with some preliminary results about sequents ${}\rvvdash{}{A,B}$ in \pOLf{} which will be the target of sequents of \OL.

\begin{lem}\label{lempolfex}
  The following rules are admissible in \pOLf:
  \begin{equation*}
    \ZIC{\rvvdash{}{C,\top}}
    \DP
\qquad
    \AIC{\rvvdash{}{C,A}}
    \AIC{\rvvdash{}{C,B}}
    \BIC{\rvvdash{}{C,A\wedge B}}
    \DP
\qquad
    \AIC{\rvvdash{A}{C}}
    \UIC{\rvvdash{}{C,A}}
    \DP
\qquad
    \AIC{\rvvdash{}{A,C}}
    \UIC{\rvvdash{}{C,A}}
    \DP
  \end{equation*}
\end{lem}

\proof
The first three rules are obtained by induction on $C$.
The fourth is obtained by induction on the proof of ${}\rvvdash{}{A,C}$, using the other three.
\qed

\begin{lem}\label{lemvee}
In \pOLf, the following rules are admissible (and similarly for $B\vee A$ instead of $A\vee B$):
\begin{equation*}
\begin{array}{c}
  \text{$A$ asynchronous}
  \\[1ex]
  \AIC{\rvvdash{C}{A}}
  \UIC{\rvvdash{A\vee B}{C}}
  \DP
\end{array}
\qquad\qquad
\begin{array}{c}
  \text{$A$ synchronous}
  \\[1ex]
  \AIC{\rvvdash{A}{C}}
  \UIC{\rvvdash{A\vee B}{C}}
  \DP
\end{array}
\qquad
\begin{array}{c}
  \text{$A$ synchronous}
  \\[1ex]
  \AIC{\rvvdash{A,C}{}}
  \UIC{\rvvdash{A\vee B,C}{}}
  \DP
\end{array}
\qquad
\begin{array}{c}
  \text{$A$ synchronous}
  \\[1ex]
  \AIC{\fcvdash{A}{C}}
  \UIC{\fcvdash{A\vee B}{C}}
  \DP
\end{array}
\end{equation*}
\end{lem}

\proof
If $A$ is asynchronous, we have:
\begin{prooftree}
  \AIC{\rvvdash{C}{A}}
  \REACfc{\fcvdash{C}{A}}
  \VEEl{\fcvdash{C}{A\vee B}}
  \DECIDl{\rvvdash{A\vee B,C}{}}
  \REACrv{\rvvdash{A\vee B}{C}}
\end{prooftree}

If $A$ is synchronous, we prove the three statements by mutual induction on the proof. Key cases are:
\begin{equation*}
  \AIC{\fcvdash{A}{A}}
  \noLine
 \renewcommand{\extraVskip}{0pt}
  \UIC{\vdots}
 \renewcommand{\extraVskip}{2pt}
  \noLine
  \UIC{\rvvdash{A,A}{}}
  \CWl{\rvvdash{A,C}{}}
  \DP
\qquad\qquad\mapsto\qquad\qquad
  \AIC{\text{IH}}
  \noLine
  \UIC{\fcvdash{A\vee B}{A}}
  \VEEl{\fcvdash{A\vee B}{A\vee B}}
  \DECIDr{\rvvdash{A\vee B,A\vee B}{}}
  \CWl{\rvvdash{A\vee B,C}{}}
  \DP
\end{equation*}
and
\begin{equation*}
  \AIC{\fcvdash{C}{A}}
  \DECIDl{\rvvdash{A,C}{}}
  \DP
\qquad\qquad\mapsto\qquad\qquad
  \AIC{\fcvdash{C}{A}}
  \VEEl{\fcvdash{C}{A\vee B}}
  \DECIDl{\rvvdash{A\vee B,C}{}}
  \DP \tag*{\raisebox{-1.25em}{\qEd}}
\end{equation*}
\def\popQED{}

\begin{prop}[Axiom expansion for \pOLf]\label{proppolfetaexp}
  If $A$ is synchronous or a negation of a variable, ${}\rvvdash{A}{\neg A}$ is provable.
\end{prop}

\proof
The proof goes by induction on the size of $A$. The main case is when $A=A_1\vee A_2$. We must consider whether each $A_i$ is synchronous or asynchronous.
If $A_i$ is asynchronous, by induction hypothesis, we have ${}\rvvdash{\neg A_i}{A_i}$ and, by Lemma~\ref{lemvee}, ${}\rvvdash{A_1\vee A_2}{\neg A_i}$.
If $A_i$ is synchronous, we have ${}\rvvdash{A_i}{\neg A_i}$ by induction hypothesis and, by Lemma~\ref{lemvee}, it implies ${}\rvvdash{A_1\vee A_2}{\neg A_i}$. We thus have ${}\rvvdash{A_1\vee A_2}{\neg A_i}$ ($i\in\{1,2\}$) in any case and we can conclude:
\begin{equation*}
  \AIC{\rvvdash{A_1\vee A_2}{\neg A_1}}
  \AIC{\rvvdash{A_1\vee A_2}{\neg A_2}}
  \WEDGErv{\rvvdash{A_1\vee A_2}{\neg A_1\wedge\neg A_2}}
  \DP \tag*{\raisebox{-0.6em}{\qEd}}
\end{equation*}
\def\popQED{}

This leads us to the completeness of \pOLf{} for orthologic by means of the completeness of \OL{} and the following translation result:

\begin{thm}[Completeness of \pOLf]\label{thmoltopolf}
  If ${}\vdash A,B$ is provable in \OL{} then ${}\rvvdash{}{A,B}$ is provable in \pOLf.
\end{thm}

\proof
By induction on the proof of ${}\vdash A,B$ in \OL, the main cases are:
\begin{itemize}
\item If the last rule is a contraction-weakening rule, we use Lemma~\ref{lemolrenv} to restrict ourselves to the ($\cw_{\vee}$) case, and by induction hypothesis we have ${}\rvvdash{}{A_1\vee A_2,A_1\vee A_2}$.
The only way this is provable is by:
\begin{prooftree}
  \AIC{\rvvdash{A_1\vee A_2,A_1\vee A_2}{}}
  \REACrv{\rvvdash{A_1\vee A_2}{A_1\vee A_2}}
  \REACrr{\rvvdash{}{A_1\vee A_2,A_1\vee A_2}}
\end{prooftree}
so that we can build:
 \begin{prooftree}
   \AIC{\rvvdash{A_1\vee A_2,A_1\vee A_2}{}}  
   \CWl{\rvvdash{A_1\vee A_2,B}{}}  
   \REACrv{\rvvdash{A_1\vee A_2}{B}}  
   \REACrr{\rvvdash{}{A_1\vee A_2,B}}
 \end{prooftree}
\item If the last rule is a ($\veel$) rule, by induction hypothesis we have ${}\rvvdash{}{A,C}$, thus using Lemmas~\ref{lempolfex} and~\ref{lemvee}, Proposition~\ref{proppolfetaexp} and Theorem~\ref{thmpolfcutelim}:
\begin{equation*}
    \AIC{\rvvdash{}{A,C}}
    \dashedLine
    \UIC{\rvvdash{}{C,A}}
    \AIC{\text{$A$ synchronous}}
    \dashedLine
    \UIC{\rvvdash{A}{\neg A}}
    \dashedLine
    \UIC{\rvvdash{A\vee B}{\neg A}}
    \dashedLine
    \CUTRB{\rvvdash{A\vee B}{C}}
    \REACrr{\rvvdash{}{A\vee B,C}}
    \DP
\qquad\text{and}\qquad
    \AIC{\rvvdash{}{A,C}}
    \dashedLine
    \UIC{\rvvdash{}{C,A}}
    \AIC{\text{$A$ asynchronous}}
    \dashedLine
    \UIC{\rvvdash{\neg A}{A}}
    \dashedLine
    \UIC{\rvvdash{A\vee B}{\neg A}}
    \dashedLine
    \CUTRB{\rvvdash{A\vee B}{C}}
    \REACrr{\rvvdash{}{A\vee B,C}}
    \DP \tag*{\raisebox{-2.6em}{\qEd}}
\end{equation*}
\def\popQED{}
\end{itemize}

As promised in Section~\ref{secol}, we can deduce cut elimination for \OL.

\begin{prop}[Cut Elimination for \OL]\label{propolcutelim}
  The cut rule is admissible in \OL.
\end{prop}

\proof
By Theorem~\ref{thmoltopolf}, we have ${}\rvvdash{}{A,B}$ and ${}\rvvdash{}{\neg B,C}$ in \pOLf. By Lemma~\ref{lempolfex} we deduce ${}\rvvdash{}{C,\neg B}$. Using $\cutra$ (Theorem~\ref{thmpolfcutelim}) we have ${}\rvvdash{}{A,C}$, and by Proposition~\ref{proppolftool}, ${}\vdash A,C$ in \OL.
\qed

\subsection{A Second Focused System \texorpdfstring{\OLf}{\OLft}}\label{secolf}

If we try to apply a simple bottom-up proof-search procedure in a sequent calculus system, a first obstacle to the finiteness of the search is given by cut rules. If a cut rule cannot be eliminated then a given conclusion leads us to a possibly infinite set of premises. A second obstacle comes from loops, \ie{} non trivial derivations leading from a sequent to the same sequent (note however this obstacle can be dealt with by using loop detection during the search, but loops make the proof-search longer).
All the systems we have seen so far contain non-trivial loops:
\begin{gather*}
\allowdisplaybreaks
  \AIC{B\vdash A}
  \ZIC{B\vdash B}
  \BIC{B\vdash A\wedge B}
  \ZIC{A\wedge B\vdash A}
  \BIC{B\vdash A}
  \DP
\qquad\qquad
  \begin{array}{c}
    \OCL \\[1ex]
    \AIC{\vdash A\vee B}
    \WKR{\vdash A,A\vee B}
    \VEERl{\vdash A\vee B}
    \DP
  \end{array}
\\[2ex]
\allowdisplaybreaks
  \begin{array}{c}
    \OL \\[1ex]
    \AIC{\vdash A\vee B,A\vee B}
    \CW{\vdash A\vee B,A}
    \EX{\vdash A,A\vee B}
    \VEEl{\vdash A\vee B,A\vee B}
    \DP
  \end{array}
\qquad\qquad
  \begin{array}{c}
    \pOLf \\[1ex]
    \AIC{\rvvdash{\neg X\vee B,\neg X\vee B}{}}
    \CWl{\rvvdash{\neg X\vee B,\neg X}{}}
    \REACrv{\rvvdash{\neg X\vee B}{\neg X}}
    \REACfc{\fcvdash{\neg X\vee B}{\neg X}}
    \VEEl{\fcvdash{\neg X\vee B}{\neg X\vee B}}
    \DECIDl{\rvvdash{\neg X\vee B,\neg X\vee B}{}}
    \DP
  \end{array}
\end{gather*}

Avoiding loops is one of the motivations for looking for a more constrained focused system.
Let us analyse loops in \pOLf. They mainly come from rules acting on sequents of the shape ${}\rvvdash{\_,\_}{}$. If we look at derivations in a bottom-up way, we reach such a sequent through a ($\reacrv$) rule:
\begin{prooftree}
  \AIC{\rvvdash{C,A}{}}
  \REACrv{\rvvdash{C}{A}}
\end{prooftree}
then we stay with sequents ${}\rvvdash{\_,\_}{}$ by using (upwardly):
\begin{equation*}
  \AIC{\rvvdash{C,C}{}}
  \CWl{\rvvdash{C,A}{}}
  \DP
\qquad \text{and} \qquad
  \AIC{\rvvdash{C,C}{}}
  \CWr{\rvvdash{A,C}{}}
  \DP
\end{equation*}
until we reach:
\begin{equation*}
  \AIC{\fcvdash{C}{A}}
  \DECIDl{\rvvdash{A,C}{}}
  \DP
\qquad \text{or} \qquad
  \AIC{\fcvdash{C}{A}}
  \DECIDr{\rvvdash{C,A}{}}
  \DP
\;.
\end{equation*}
Globally, this means we start with a sequent ${}\rvvdash{C}{A}$ and we must end with ${}\fcvdash{C}{A}$, ${}\fcvdash{A}{C}$, ${}\fcvdash{A}{A}$ or ${}\fcvdash{C}{C}$. This would correspond to four derivable rules:
\begin{equation*}
  \AIC{\fcvdash{C}{A}}
  \dashedLine
  \UIC{\rvvdash{C}{A}}
  \DP
\qquad\qquad
  \AIC{\fcvdash{A}{C}}
  \dashedLine
  \UIC{\rvvdash{C}{A}}
  \DP
\qquad\qquad
  \AIC{\fcvdash{A}{A}}
  \dashedLine
  \UIC{\rvvdash{C}{A}}
  \DP
\qquad\qquad
  \AIC{\fcvdash{C}{C}}
  \dashedLine
  \UIC{\rvvdash{C}{A}}
  \DP
\end{equation*}
In the same time we want to try to constrain contraction so that it is applied on $\vee$-formulas only (in the spirit of Lemma~\ref{lemolrenv}). Moreover we would like contraction not being applied twice on the same formula. In particular we get read of the fourth rule just above, which would allow $C$ to be contracted (uselessly) many times.
All these remarks lead us to the following new focused system called \OLf:
\boxit{\OLf}{
\begin{gather*}
  \AIC{\rvvdash{}{A,C}}
  \AIC{\rvvdash{}{B,C}}
  \WEDGErr{\rvvdash{}{A\wedge B,C}}
  \DP
\qquad\qquad
  \TOPrr{\rvvdash{}{\top,C}}
  \DP
\\[2ex]
  \AIC{\rvvdash{C}{A}}
  \AIC{\rvvdash{C}{B}}
  \WEDGErv{\rvvdash{C}{A\wedge B}}
  \DP
\qquad\qquad
  \SC{\text{(s) or (n)}}
  \TOPrv{\rvvdash{A}{\top}}
  \DP  
\\[2ex]
  \AIC{\fcvdash{B\vee C}{B\vee C}}
  \CWrr{\rvvdash{}{B\vee C,A}}
  \DP
\qquad\qquad
  \AIC{\fcvdash{B\vee C}{B\vee C}}
  \SC{\text{(s) or (n)}}
  \CWrv{\rvvdash{A}{B\vee C}}
  \DP
\\[2ex]
  \AXX{\fcvdash{\neg X}{X}}
  \DP
\qquad\qquad
  \AIC{\fcvdash{C}{A}}
  \VEEl{\fcvdash{C}{A\vee B}}
  \DP
\qquad\qquad
  \AIC{\fcvdash{C}{B}}
  \VEEr{\fcvdash{C}{A\vee B}}
  \DP
\\[2ex]
  \AIC{\rvvdash{A}{C}}
  \REACrr{\rvvdash{}{A,C}}
  \DP
\qquad
  \AIC{\rvvdash{C}{A}}
  \SC{\text{(a)}}
  \REACfc{\fcvdash{C}{A}}
  \DP
\qquad
  \AIC{\fcvdash{C}{A}}
  \SC{\text{(s)}}
  \DECIDl{\rvvdash{A}{C}}
  \DP
\qquad
  \AIC{\fcvdash{C}{A}}
  \SC{\text{(s)}}
  \DECIDr{\rvvdash{C}{A}}
  \DP
\end{gather*}
\hfil
$\text{(a) $A$ is asynchronous} \quad
\text{(s) $A$ is synchronous} \quad
\text{(n) $A$ is the negation of a variable}$
}

Note, sequents ${}\rvvdash{A,B}{}$ disappear in this system which relies on three kinds of sequents only: ${}\rvvdash{A}{B}$, ${}\fcvdash{A}{B}$ and ${}\rvvdash{}{A,B}$.

\begin{exa}
  We can prove in \OLf{} the sequent associated with Example~\ref{exone}:
  \begin{prooftree}
    \AXX{\fcvdash{\neg X}{X}}
    \DECIDl{\rvvdash{X}{\neg X}}
    \REACfc{\fcvdash{X}{\neg X}}
    \VEEr{\fcvdash{X}{(X\wedge Y)\vee \neg X}}
    \VEEl{\fcvdash{X}{((X\wedge Y)\vee \neg X)\vee \neg Y}}
    \DECIDl{\rvvdash{((X\wedge Y)\vee \neg X)\vee \neg Y}{X}}
    \AXX{\fcvdash{\neg Y}{Y}}
    \DECIDl{\rvvdash{Y}{\neg Y}}
    \REACfc{\fcvdash{Y}{\neg Y}}
    \VEEr{\fcvdash{Y}{((X\wedge Y)\vee \neg X)\vee \neg Y}}
    \DECIDl{\rvvdash{((X\wedge Y)\vee \neg X)\vee \neg Y}{Y}}
    \WEDGErv{\rvvdash{((X\wedge Y)\vee \neg X)\vee \neg Y}{X\wedge Y}}
    \REACfc{\fcvdash{((X\wedge Y)\vee \neg X)\vee \neg Y}{X\wedge Y}}
    \VEEl{\fcvdash{((X\wedge Y)\vee \neg X)\vee \neg Y}{(X\wedge Y)\vee \neg X}}
    \VEEl{\fcvdash{((X\wedge Y)\vee \neg X)\vee \neg Y}{((X\wedge Y)\vee \neg X)\vee \neg Y}}
    \CWrv{\rvvdash{\bot}{((X\wedge Y)\vee \neg X)\vee \neg Y}}
    \REACrr{\rvvdash{}{\bot,((X\wedge Y)\vee \neg X)\vee \neg Y}}
  \end{prooftree}
\end{exa}

The system \OLf{} is as expressive as \pOLf{} for sequents ${}\rvvdash{}{A,B}$. In particular:
\begin{prop}[Expressiveness of \OLf]\label{propolfcmp}
  If ${}\rvvdash{}{A,B}$ is provable in \pOLf{}, it is also provable in \OLf.
\end{prop}

\proof
We prove by induction on the proof $\pi$ in \pOLf{} the more general statement:
\begin{itemize}
\item If ${}\rvvdash{}{A,B}$ in \pOLf{} then ${}\rvvdash{}{A,B}$ in \OLf.
\item If ${}\rvvdash{A}{B}$ in \pOLf{} then either ${}\rvvdash{A}{B}$ in \OLf{} or $A=A_1\vee A_2$ with ${}\fcvdash{A}{A}$ in \OLf.
\item If ${}\fcvdash{A}{B}$ in \pOLf{} then either ${}\fcvdash{A}{B}$ in \OLf{} or $A=A_1\vee A_2$ with ${}\fcvdash{A}{A}$ in \OLf.
\item If ${}\rvvdash{A,B}{}$ in \pOLf{} then at least one of the following four possibilities holds:
  \begin{itemize}
  \item $B$ is synchronous and ${}\fcvdash{A}{B}$ in \OLf;
  \item $A$ is synchronous and ${}\fcvdash{B}{A}$ in \OLf;
  \item $A=A_1\vee A_2$ and ${}\fcvdash{A}{A}$ in \OLf;
  \item $B=B_1\vee B_2$ and ${}\fcvdash{B}{B}$ in \OLf.
  \end{itemize}
\end{itemize}
We consider each possible last rule for $\pi$. Interesting cases are:
\begin{itemize}
\item For the two contraction rules, we have ${}\rvvdash{C,C}{}$ in \pOLf{} thus, by induction hypothesis, ${}\fcvdash{C}{C}$ in \OLf{} with $C$ synchronous and we are done since ${}\fcvdash{\bot}{\bot}$ and ${}\fcvdash{X}{X}$ are not provable thus $C=C_1\vee C_2$.
\item For $(\veel)$, by induction hypothesis, we have either ${}\fcvdash{C}{A}$ or ${}\fcvdash{C_1\vee C_2}{C_1\vee C_2}$ in \OLf{} with $C=C_1\vee C_2$. In the first case, we apply the corresponding rule. In the second case, we are immediately done.
\item For $(\reacrr)$, by induction hypothesis we have ${\!}\rvvdash{\!A}{C}$
  or $A=A_1\vee A_2$ with \hbox{${\!}\fcvdash{\!A_1\!\vee\! A_2}{A_1\!\vee\!A_2}$}, we can build:
  \begin{center}
    \AIC{\rvvdash{A}{C}}
    \REACrr{\rvvdash{}{A,C}}
    \DP
    \qquad or\qquad
    \AIC{\fcvdash{A_1\vee A_2}{A_1\vee A_2}}
    \CWrr{\rvvdash{}{A_1\vee A_2,C}}
    \DP
  \end{center}
\item For $(\reacrv)$, we apply the induction hypothesis and we obtain four possible cases:
  \begin{itemize}
  \item If ${}\fcvdash{C}{A}$ in \OLf{} with $A$ synchronous, we have:
      \AIC{{}\fcvdash{C}{A}}
      \DECIDr{{}\rvvdash{C}{A}}
      \DP
  \item If ${}\fcvdash{A}{C}$ in \OLf{} with $C$ synchronous, we have:
      \AIC{{}\fcvdash{A}{C}}
      \DECIDl{{}\rvvdash{C}{A}}
      \DP
  \item If ${}\fcvdash{C_1\vee C_2}{C_1\vee C_2}$ ($C=C_1\vee C_2$) in \OLf, we are done.
  \item If ${}\fcvdash{A_1\vee A_2}{A_1\vee A_2}$ ($A=A_1\vee A_2$) in \OLf, we have:
    \begin{equation*}
      \AIC{{}\fcvdash{A_1\vee A_2}{A_1\vee A_2}}
      \CWrv{{}\rvvdash{C}{A_1\vee A_2}}
      \DP \tag*{\raisebox{-.6em}{\qEd}}
    \end{equation*}
    \def\popQED{}
  \end{itemize}
\end{itemize}

\begin{prop}[Soundness of \OLf]\label{propolfsnd}
  If ${}\rvvdash{}{A,B}$ is provable in \OLf{} then ${}\vdash A,B$ is provable in \OL.
\end{prop}

\proof
  Similar to the proof of Proposition~\ref{proppolftool}.
\qed

From Propositions~\ref{propolcutelim},~\ref{propolfcmp} and~\ref{propolfsnd}, and Theorem~\ref{thmoltopolf}, we can deduce the admissibility of the following cut rule in \OLf:
\begin{prooftree}
  \AIC{{}\rvvdash{}{A,B}}
  \AIC{{}\rvvdash{}{\neg B,C}}
  \CUT{{}\rvvdash{}{A,C}}
\end{prooftree}

We have thus built yet another sound and complete system for orthologic.
This one has stronger constraints on the structure of proofs than the previous ones.
A key property of this new system (which holds in none of the previous ones) is the termination of the naive bottom-up proof search strategy (Proposition~\ref{propfinbranch}). 

\section{Proof Search in \texorpdfstring{\OLf}{\OLft}}\label{secps}

We first develop a few properties of \OLf{} on which we will rely for proof search. In a second time, we will compare with other algorithms from the literature.

\subsection{Backward Proof Search}\label{secbwps}

The basic idea of backward proof search in a cut-free sequent calculus system is to start from the sequent to be proved, to look in a bottom-up manner at each possible instance of a rule with this sequent as conclusion and to continue recursively with the premises of these instances until axioms are reached.
Given a sequent, we are going to bound the length of branches of its proofs in \OLf, thus proving the termination of this algorithm.
Let us first define the following measure on formulas:
\begin{align*}
  \fbound{X} = \fbound{\neg X} &= \fbound{\bot} = \fbound{\top} = 1 \\
  \fbound{A\wedge B} &= \fbound{A} + \fbound{B} \\
  \fbound{A\vee B} &= 2\fbound{A} + 2\fbound{B}
\end{align*}
As a bound on $\fboundsym$, we have $\fbound{A} < 2^{\size{A}}$ where $\size{A}$ is the size (number of symbols) of $A$.

\begin{prop}[Finiteness of Branches in \OLf]\label{propfinbranch}
  Given two formulas $A$ and $B$, $2\fbound{A}+2\fbound{B}$ is a bound on the length of the branches of any proof of ${}\rvvdash{}{A,B}$ in \OLf.
\end{prop}

\proof
We define the measure $\sboundsym$ of a sequent, according to its shape:
\begin{align*}
  \sbound{{}\rvvdash{}{A,B}} &= 2\fbound{A}+2\fbound{B} \\
  \sbound{{}\rvvdash{A}{B}}  &= \fbound{A}+2\fbound{B} \\
  \sbound{{}\fcvdash{A}{B}}  &= 
    \begin{cases}
      \fbound{A}+\fbound{B} & \text{if $B$ is synchronous} \\
      \fbound{A}+2\fbound{B}+1 & \text{if $B$ is asynchronous}
    \end{cases}
\end{align*}
We now prove for each rule of \OLf: if $S_1$ is a sequent premise of the rule and $S_2$ is the sequent conclusion of the rule, then $\sbound{S_1} < \sbound{S_2}$.
For example:
\begin{center}
\begin{tabular}{l@{\qquad\quad}l}
\qquad ($\veel$) with $A$ synchronous & \qquad ($\veel$) with $A$ asynchronous \\[1ex]
$\begin{array}{r@{\;}c@{\;}l}
  \sbound{{}\fcvdash{C}{A}} &=& \fbound{C}+\fbound{A} \\
  &<& \fbound{C}+2\fbound{A} + 2\fbound{B} \\
  &=& \fbound{C}+\fbound{A\vee B} \\
  &=& \sbound{{}\fcvdash{C}{A\vee B}}
\end{array}$ &
$\begin{array}{r@{\;}c@{\;}l}
  \sbound{{}\fcvdash{C}{A}} &=& \fbound{C}+2\fbound{A}+1 \\
  &<& \fbound{C}+2\fbound{A} + 2\fbound{B} \\
  &=& \fbound{C}+\fbound{A\vee B} \\
  &=& \sbound{{}\fcvdash{C}{A\vee B}}
\end{array}$
\end{tabular}
\end{center}
Thus for any sequent $S$, $\sbound{S}$ is a bound on the height of the branches of the proofs of $S$.
\qed

Since rules of \OLf{} are finitely branching, this bound on the length of branches ensures (the absence of loops and) the termination of the backward proof search.
Moreover, thanks to the sub-formula property, we know every sequent appearing in a proof of ${}\rvvdash{}{A,B}$ is made of two formulas which are sub-formulas of $A$ or $B$. Since we have three different kinds of sequents, there are at most $3(\size{A}+\size{B})^2$ such sequents. We have just proved a sequent cannot appear twice in a branch of a proof, so we can deduce a tighter bound than $\sboundsym$ on the height of branches: $3(\size{A}+\size{B})^2$. We thus have an upper bound $2^{3(\size{A}+\size{B})^2+1}$ on the size of proofs since rules have arity at most $2$.

\subsection{Single Formula Proof Search}\label{secdiag}

As we have seen in Section~\ref{secol}, in systems with exactly two formulas in sequents presented in this paper, the provability of a formula $A$ in orthologic is encoded as the provability of a sequent of the shape ${}\vdash A,A$ or ${}\rvvdash{}{A,A}$. Since we are often interested in the provability of a single formula, these sequents play a specific role.
Note however that there is no direct reduction of the provability of two-formulas sequents to the provability of these single-formula sequents as shown by the following proposition. This is related with the so-called ``implication problem'' in quantum logic (see~\cite{omlogic} for example).

\begin{prop}[Single-Formula Provability]\label{propimppb}
   In \OL, there is no formula $F$ using only two variables $X$ and $Y$ and such that, for any two formulas $A$ and $B$:
   \begin{equation*}
{}\vdash A,B\iff{}\vdash\bisubst{F}{A}{X}{B}{Y},\bisubst{F}{A}{X}{B}{Y}
\end{equation*}
\end{prop}

\proof
First note that ${}\vdash C,C$ if and only if ${}\vdash C,\bot$.
Now, if $F$ exists, it defines a function $I:\FOL\times\FOL\rightarrow\FOL$ (where $\FOL$ is the free ortholattice over the set $\Var$) by mapping $(p,q)$ to $I(p,q)=\bisubst{F}{\neg p}{X}{q}{Y}$ (variables are mapped to their corresponding element in $\FOL$ and conjunction, disjunction and negation are interpreted by the corresponding ortholattice operations). We then have:
\begin{align*}
  p\leq q\text{ in $\FOL$} &\iff {}\vdash\neg p,q\text{ in \OL} \\
  &\iff {}\vdash \bisubst{F}{\neg p}{X}{q}{Y},\bot\text{ in \OL} \\
  &\iff \top\leq \bisubst{F}{\neg p}{X}{q}{Y}\text{ in \FOL} \\
  &\iff I(p,q)=\top\text{ in \FOL}
\end{align*}
which contradicts the main theorem of~\cite{wmrequired}, since $\FOL$ is not orthomodular (see Example~\ref{exhexa}).
\qed

Nevertheless since the restricted sequents ${}\vdash A,A$ or ${}\rvvdash{}{A,A}$ arise naturally (in particular when considering provability of formulas instead of sequents), we give a dedicated look at them.
We can give some optimisation on the bottom structure of proofs of sequents ${}\rvvdash{}{A,A}$ in \OLf.

\begin{prop}[Diagonal Sequent]\label{propdiag}
  The following properties hold in \OLf:
  \begin{itemize}
  \item ${}\rvvdash{}{X,X}$, ${}\rvvdash{}{\neg X,\neg X}$ and ${}\rvvdash{}{\bot,\bot}$ are not provable (for any $X$).
  \item ${}\rvvdash{}{\top,\top}$ is provable.
  \item ${}\rvvdash{}{B\wedge C,B\wedge C}$ is provable if and only if both ${}\rvvdash{}{B,B}$ and ${}\rvvdash{}{C,C}$ are provable.
  \item ${}\rvvdash{}{B\vee C,B\vee C}$ is provable if and only if ${}\fcvdash{B\vee C}{B\vee C}$ is provable.
  \end{itemize}
\end{prop}

\proof
For the first two points, we can for example move to \OL{} thanks to Proposition~\ref{propolfsnd}.
For $\wedge$, we move back and forth to \OL{} thanks to Theorem~\ref{thmoltopolf} and Propositions~\ref{propolfcmp} and~\ref{propolfsnd}. In \OL, we use Lemma~\ref{lemolrevw} and:
\begin{prooftree}
  \AIC{\vdash B,B}
  \CW{\vdash B,B\wedge C}
  \AIC{\vdash C,C}
  \CW{\vdash C,B\wedge C}
  \WEDGE{\vdash B\wedge C,B\wedge C}
\end{prooftree}
For $\vee$, the only possible last rules are:
\begin{equation*}
  \AIC{\fcvdash{B\vee C}{B\vee C}}
  \CWrr{\rvvdash{}{B\vee C,B\vee C}}
  \DP
\qquad\qquad
  \AIC{\rvvdash{B\vee C}{B\vee C}}
  \REACrr{\rvvdash{}{B\vee C,B\vee C}}
  \DP
\end{equation*}
and for a proof of ${}\rvvdash{B\vee C}{B\vee C}$, the only possible last rules are:
\begin{gather*}
  \AIC{\fcvdash{B\vee C}{B\vee C}}
  \CWrv{\rvvdash{B\vee C}{B\vee C}}
  \DP
\qquad\qquad
  \AIC{\fcvdash{B\vee C}{B\vee C}}
  \DECIDl{\rvvdash{B\vee C}{B\vee C}}
  \DP
\qquad\qquad
  \AIC{\fcvdash{B\vee C}{B\vee C}}
  \DECIDr{\rvvdash{B\vee C}{B\vee C}}
  \DP
\end{gather*}
so that ${}\fcvdash{B\vee C}{B\vee C}$ must be provable for ${}\rvvdash{}{B\vee C,B\vee C}$ to be provable.
In the other direction we directly use ($\cwrr$).
\qed

This means in particular that any sequent ${}\vdash A,A$ is either clearly not provable or equivalent to a finite family of sequents ${}\fcvdash{B_1\vee C_1}{B_1\vee C_1}$,\dots, ${}\fcvdash{B_n\vee C_n}{B_n\vee C_n}$ (with each $B_i\vee C_i$ sub-formula of $A$).

\subsection{Forward Proof Search}\label{secfwps}

Forward proof search consists in building, in a top-down way, proof-trees which are candidates to be sub-proof-trees of proofs of a given sequent.
Clearly the sub-formula property can be used to control the sequents to be considered inside the proof-trees. We can use even stronger constraints based on the structure of focused sequents (see Lemma~\ref{lemsidecond} for example).

Following Section~\ref{secdiag}, we present some specific optimisations which can be applied in the case of forward proof search for sequents ${}\rvvdash{}{A,A}$.
Let us fix a formula $A$. We want to study sub-proof-trees of proofs of ${}\rvvdash{}{A,A}$ in \OLf. Thanks to Proposition~\ref{propdiag}, it is then enough to focus on proofs of sequents of the shape ${}\fcvdash{B\vee C}{B\vee C}$.

\begin{prop}[Strengthened Sub-Formula Property]\label{propfs}
  If ${}\fcvdash{D}{E}$ or ${}\rvvdash{D}{F}$ appears in a proof of ${}\fcvdash{B\vee C}{B\vee C}$ in \OLf, $D$, $E$ and $F$ are sub-formulas of $B\vee C$ and moreover:
  \begin{itemize}
  \item if $D$ is synchronous, it is equal to $B\vee C$ or it appears inside $B\vee C$ just below a $\wedge$ connective;
  \item if $E$ is asynchronous, it appears inside $B\vee C$ just below a $\vee$ connective;
  \item if $F$ is synchronous, it appears inside $B\vee C$ just below a $\wedge$ connective.
  \end{itemize}
\end{prop}

\proof
  Since ${}\fcvdash{B\vee C}{B\vee C}$ satisfies the conclusion of the statement, we prove for each rule that if the conclusion satisfies it, then all its premises as well.
  \begin{itemize}
  \item For the $\wedgerv$ rule, we apply the hypothesis relative to position $D$ to the conclusion and, concerning $F$, we have formulas occurring below a $\wedge$ connective.
  \item For the $\cwrv$ rule, position $D$ in the premise is below a $\wedge$ connective since it is the case for position $F$ in the conclusion.
  \item For the $\vee$ rules, the result is immediate.
  \item For the $\reacfc$ rule, the formula in position $F$ in the premise must be asynchronous.
  \item For the $\decidl$ rule, if it is synchronous, position $D$ in the premise is below a $\wedge$ connective since it is the case for position $F$ in the conclusion.
  \item For the $\decidr$ rule, it is immediate from the property of the conclusion and the side condition. \qedhere
  \end{itemize}

This proposition provides us constraints on the meaningful sequents to be considered during forward proof search for sequents ${}\rvvdash{}{A,A}$. This means we can restrict the application of rules in the algorithm for forward proof search to the case where they generate sequents satisfying the properties given by Proposition~\ref{propfs}.

\subsection{Benchmark}\label{secbench}

We want to compare our proof-search procedures with procedures from the literature.
We consider some formulas from~\cite{exortho} and~\cite{proofsearchortho} as well as some random formulas in the language of orthologic:
\begin{align*}
\cunea
       & = ((\neg X\vee Y)\wedge X)
           \; \vee \;
	       ( (X\wedge\neg Y)
               \; \vee \;
		 ( (\neg X\wedge((X\vee\neg Y)\wedge (X\vee Y))) \\
                 &\qquad\qquad\qquad \; \vee \;
		   (\neg X\wedge ((\neg X\wedge Y)\vee(\neg X\wedge \neg Y)))))
\\
  \cuneb &= X\; \vee \; ((\neg X\wedge ((X\vee\neg Y)\wedge (X\vee Y)))
         \; \vee \;
	 (\neg X\wedge ((\neg X\wedge Y)\vee (\neg X\wedge \neg Y)))) \\
  \cunec
&=  (((X\vee \neg Y)\wedge (X\vee Y))\wedge (\neg X\vee (X\wedge\neg Y))) \; \vee \; (\neg X\vee Y) \\
  \Phi_0 &= X_0\vee\neg X_0 \qquad
  \Phi_{n+1} =   ((X_n\wedge Y_n)\wedge(X_n\wedge Z_n)) \vee
              (((\neg X_n\wedge \Phi_n)\vee\neg Y_n)\vee\neg Z_n) \\
  \Psi^1_0 & = \top \qquad \Psi^2_0 = \bot \qquad
  \Psi^1_{n+1} = \Psi^1_n\wedge X_n \qquad
  \Psi^2_{n+1} = \Psi^2_n\vee Y_n
 \\
  \Psi^3_n &=  (X\vee (Y\wedge \Psi^2_n))\wedge \Psi^1_n \qquad
  \Psi^4_n = (Y\wedge (X\vee \Psi^1_n))\vee \Psi^2_n \qquad
  \Psi_n = \neg \Psi^3_n\vee \Psi^4_n
\end{align*}
The formulas $\cuneb$, $\cunec$ and $\Phi_n$ are provable, while $\cunea$ and $\Psi_n$ are not.

We compare four algorithms:
\begin{itemize}
\item \cf{} is \logsys{prove-cf}~\cite{proofsearchortho}, a mixed backward-forward algorithm (pure backward search would loop in general);
\item \fw{} is the forward algorithm from~\cite{proofsearchortho};
\item \bwf{} is the backward algorithm based on \OLf{} (see Section~\ref{secbwps});
\item \fwf{} is the forward algorithm based on \OLf{} using Proposition~\ref{propfs} (see Section~\ref{secfwps}).
\end{itemize}
The implementations are done in \ocaml{} in the most naive way (except that we use some memoization), so that running time (\emph{time}, in seconds) should not be taken too seriously. As an alternative measure which depends less on the particular implementation, we also count the number of rule occurrences (\emph{number of rules}) applied during search.

\begin{center}
\begin{tabular}{|l!{\vrule width 2pt}r|r||r|r!{\vrule width 2pt}r|r||r|r!{\vrule width 2pt}}
\hline
 & \multicolumn{4}{c!{\vrule width 2pt}}{\emph{time} (in seconds)} & \multicolumn{4}{c!{\vrule width 2pt}}{\emph{number of rules}} \\
\hline
 & \multicolumn{1}{c|}{\cf} & \multicolumn{1}{c||}{\bwf} & \multicolumn{1}{c|}{\fw} & \multicolumn{1}{c!{\vrule width 2pt}}{\fwf} & \multicolumn{1}{c|}{\cf} & \multicolumn{1}{c||}{\bwf} & \multicolumn{1}{c|}{\fw} & \multicolumn{1}{c!{\vrule width 2pt}}{\fwf} \\
\hline
$\cunea$ & 0.00 & 0.00 & 0.00 & 0.01 & 2 449 & 127 & 47 & 64 \\
\hline
$\cuneb$ & 0.00 & 0.00 & 0.00 & 0.00 & 145 & 102 & 33 & 42 \\
\hline
$\cunec$ & 0.00 & 0.00 & 0.00 & 0.00 & 78 & 142 & 43 & 49 \\
\hline
\hline
$\Phi_5$ & 0.02 & 0.00 & 0.36 & 0.28 & 5 599 & 382 & 416 & 338 \\
\hline
$\Phi_{10}$ & 0.06 & 0.00 & 6.42 & 3.37 & 11 194 & 772 & 1 791 & 1 023 \\
\hline
$\Phi_{20}$ & 0.16 & 0.00 & 120.36 & 48.36 & 22 384 & 1 552 & 7 391 & 3 443 \\
\hline
\hline
$\Psi_{5}$ & 0.18 & 0.00 & 0.17 & 0.02 & 248 855 & 303 & 343 & 123 \\
\hline
$\Psi_{10}$ & 135.86 & 0.00 & 1.06 & 0.05 & 195 724 597 & 818 & 773 & 273 \\
\hline
$\Psi_{20}$ & \timo & 0.00 & 10.21 & 0.20 & \timo & 2 598 & 2 083 & 723 \\
\hline
$\Psi_{100}$ & \timo & 0.09 & 6701.08 & 7.92 & \timo & 52 838 & 34 163 & 11 523 \\
\hline
\hline
Rnd$_{20}$ & 0.00 & 0.00 & 0.02 & 0.00 & 1 853 & 38 & 152 & 36 \\
\hline
Rnd$_{100}$ & 3.81 & 0.00 & 5.07 & 0.52 & 1 047 118 & 233 & 2 553 & 566 \\
\hline
\end{tabular}
\end{center}
Lines Rnd$_{20}$ and Rnd$_{100}$ correspond to average values over some randomly generated formulas of size respectively 20 and 100. The unknown values $\timo$ (time out) correspond to the algorithm being stopped after more than 12 hours running time without answer.

This is a minimalist benchmarking. The goal here is simply to show that the new focusing-based algorithms (\bwf{} and \fwf) look really competitive with respect to previous work (\cf{} and \fw~\cite{proofsearchortho}). For this, it is meaningful to compare \cf{} and \bwf{} on the one side, and \fw{} and \fwf{} on the other side. Comparing backward and forward algorithms is more difficult here since the time measures really depend on the specific implementation, and forward algorithms require more management of data. There is certainly room for a better implementation of data structures.

We do not answer here to the question of what would be the best proof-search algorithm for orthologic. We think future investigations in this direction should go towards focused-based approaches. While the focused backward approach looks the most promising one, the fact that \fwf{} sometimes gives better results that \bwf{} regarding the number of applied rules suggests that better implementations of \fwf{} (with more care on the chosen data structures) might become competitive.

\section{Conclusion}

We have presented new sequent-calculus proof-systems for orthologic, mainly: \OL{} which is the simplest such system we know, and \OLf{} which is based on focusing to constrain the structure of proofs. With some complementary analysis on the structure of proofs in \OLf{} we have proposed efficient proof search algorithms for orthologic which look quicker than the state of the art~\cite{proofsearchortho} (but additional studies in this direction must be done to obtain fully convincing evaluations).

Our new systems open the door for additional proof-theoretical studies of orthologic (and the possibility of extracting (finite) counter-models from proof-search failures should be investigated). We also hope this will lead to results in the theory of ortholattices (free ones in particular) in the spirit of P.~Whitman's work~\cite{freelattices}.
We plan also to work on the application of focusing to other lattice-related logics~\cite{cutelimlattices}.

We have focused on the propositional part of the logic because of its close relation with lattice theory.
From the proof theoretical point of view, it would be natural to consider first-order quantifiers as well. The theory of focusing for quantifiers is well understood and a simple extension of the systems we have considered, with $\wedge$-like rules for $\forall$ and $\vee$-like rules for $\exists$, would be easy to define.
Second-order quantification could also be investigated on the logic side in relation with complete ortholattices.

Finally, the proof theory of orthologic seems to be mature enough to try to develop some Curry-Howard correspondence aiming at exhibiting the computational content of orthologic. In particular it would be interesting to investigate notions of proof-nets for orthologic relying on the work on additive linear logic~\cite{allpn}.

\subsection*{Additional Material}\label{pageaddmat}

A \coq{} development formalising the main proofs of the paper is available at:
\begin{center}
  \url{https://arxiv.org/src/1612.01728/anc/olf_long.v}
\end{center}
The \ocaml{} code for the benchmark of Section~\ref{secbench} is available at:
\begin{center}
  \url{https://arxiv.org/src/1612.01728/anc/olf_long.ml}
\end{center}

\subsection*{Acknowledgements}

We would like to thank D.~Pous and P.~Clairambault for helpful discussions during the development of this work, as well as the anonymous referees for their comments.

\bibliographystyle{alpha}
\bibliography{olf}

\end{document}